\documentclass[12pt,draftclsnofoot,onecolumn]{./Components/MetaFiles/IEEEtran}

\usepackage{times,latexsym,psfrag,eufrak,dsfont}
\usepackage{epstopdf}
\usepackage{bbm}
\usepackage{siunitx}

\usepackage{gensymb}
\usepackage{amsmath,amssymb,amsfonts,amsthm}
\usepackage{epsfig}
\usepackage{cite}
\usepackage{hhline}
\usepackage{multirow}
\usepackage{xcolor}
\usepackage{makecell}
\usepackage{subcaption}
\usepackage{capt-of}
\usepackage[labelformat=simple]{subcaption}

\usepackage{enumerate}
\usepackage{enumitem}
\usepackage{color}
\usepackage{graphicx}
\usepackage{algorithmic}
\usepackage[ruled]{algorithm}
\usepackage{booktabs}
\usepackage{bm}
\usepackage[nolist,printonlyused]{acronym} 
\usepackage{colortbl}
\usepackage{tabularx}
\usepackage{float}
\newcolumntype{Y}{>{\centering\arraybackslash}X}

\usepackage[T1]{fontenc}
\usepackage[utf8]{inputenc}
\usepackage{comment} 
\usepackage[author={Hossein}]{pdfcomment}

\usepackage{pgf}
\usepackage{pgfplots}
\usepackage{tikz}
\usepackage{grffile}
\pgfplotsset{compat=newest}
\usetikzlibrary{plotmarks}
\usepgfplotslibrary{patchplots}
\usetikzlibrary{automata,arrows,positioning,calc,matrix,decorations.markings,decorations.pathreplacing,patterns,arrows.meta}
 \pgfdeclarepatternformonly{south west lines}{\pgfqpoint{-0pt}{-0pt}}{\pgfqpoint{3pt}{3pt}}{\pgfqpoint{3pt}{3pt}}{
        \pgfsetlinewidth{0.4pt}
        \pgfpathmoveto{\pgfqpoint{0pt}{0pt}}
        \pgfpathlineto{\pgfqpoint{3pt}{3pt}}
        \pgfpathmoveto{\pgfqpoint{2.8pt}{-.2pt}}
        \pgfpathlineto{\pgfqpoint{3.2pt}{.2pt}}
        \pgfpathmoveto{\pgfqpoint{-.2pt}{2.8pt}}
        \pgfpathlineto{\pgfqpoint{.2pt}{3.2pt}}
        \pgfusepath{stroke}}
\makeatletter
\newcommand{\vast}{\bBigg@{3}}
\newcommand{\Vast}{\bBigg@{4}}
\makeatother
\renewcommand{\IEEEQED}{\IEEEQEDopen}

\showoutput
\IEEEoverridecommandlockouts

\newtheorem{prop}{Proposition}
\newtheorem{lem}{Lemma}

\newtheorem{defi}{Definition}

\newtheorem{assumption}{Assumption}

\def\({\left(}
\def\){\right)}

\def\[{\left[}
\def\]{\right]}

\newcommand{\be}{\begin{equation}}
\newcommand{\ee}{\end{equation}}
\newcommand{\ba}{\begin{array}}
\newcommand{\ea}{\end{array}}
\newcommand{\bea}{\begin{eqnarray}}
\newcommand{\eea}{\end{eqnarray}}

\newcommand{\vbar}{\raisebox{.17ex}{\rule{.04em}{1.35ex}}}
\newcommand{\vbarind}{\raisebox{.01ex}{\rule{.04em}{1.1ex}}}
\newcommand{\D}{\ifmmode {\rm I}\hspace{-.2em}{\rm D} \else ${\rm I}\hspace{-.2em}{\rm D}$ \fi}
\newcommand{\T}{\ifmmode {\rm I}\hspace{-.2em}{\rm T} \else ${\rm I}\hspace{-.2em}{\rm T}$ \fi}
\newcommand{\B}{\ifmmode {\rm I}\hspace{-.2em}{\rm B} \else \mbox{${\rm I}\hspace{-.2em}{\rm B}$} \fi}
\newcommand{\Hil}{\ifmmode {\rm I}\hspace{-.2em}{\rm H} \else \mbox{${\rm I}\hspace{-.2em}{\rm H}$} \fi}
\newcommand{\Cind}{\ifmmode \hspace{.2em}\vbarind\hspace{-.25em}{\rm C} \else \mbox{$\hspace{.2em}\vbarind\hspace{-.25em}{\rm C}$} \fi}
\newcommand{\Q}{\ifmmode \hspace{.2em}\vbar\hspace{-.31em}{\rm Q} \else \mbox{$\hspace{.2em}\vbar\hspace{-.31em}{\rm Q}$} \fi}
\newcommand{\Z}{\ifmmode {\rm Z}\hspace{-.28em}{\rm Z} \else ${\rm Z}\hspace{-.38em}{\rm Z}$ \fi}

\newcommand{\bv}{\mbox {\boldmath $v$}}
\newcommand{\bw}{\mbox {\boldmath $w$}}

\makeatletter
\let\save@mathaccent\mathaccent
\newcommand*\if@single[3]{%
  \setbox0\hbox{${\mathaccent"0362{#1}}^H$}%
  \setbox2\hbox{${\mathaccent"0362{\kern0pt#1}}^H$}%
  \ifdim\ht0=\ht2 #3\else #2\fi
  }
\newcommand*\rel@kern[1]{\kern#1\dimexpr\macc@kerna}
\newcommand*\widebar[1]{\@ifnextchar^{{\wide@bar{#1}{0}}}{\wide@bar{#1}{1}}}
\newcommand*\wide@bar[2]{\if@single{#1}{\wide@bar@{#1}{#2}{1}}{\wide@bar@{#1}{#2}{2}}}
\newcommand*\wide@bar@[3]{%
  \begingroup
  \def\mathaccent##1##2{%
    \let\mathaccent\save@mathaccent
    \if#32 \let\macc@nucleus\first@char \fi
    \setbox\z@\hbox{$\macc@style{\macc@nucleus}_{}$}%
    \setbox\tw@\hbox{$\macc@style{\macc@nucleus}{}_{}$}%
    \dimen@\wd\tw@
    \advance\dimen@-\wd\z@
    \divide\dimen@ 3
    \@tempdima\wd\tw@
    \advance\@tempdima-\scriptspace
    \divide\@tempdima 10
    \advance\dimen@-\@tempdima
    \ifdim\dimen@>\z@ \dimen@0pt\fi
    \rel@kern{0.6}\kern-\dimen@
    \if#31
      \overline{\rel@kern{-0.6}\kern\dimen@\macc@nucleus\rel@kern{0.4}\kern\dimen@}%
      \advance\dimen@0.4\dimexpr\macc@kerna
      \let\final@kern#2%
      \ifdim\dimen@<\z@ \let\final@kern1\fi
      \if\final@kern1 \kern-\dimen@\fi
    \else
      \overline{\rel@kern{-0.6}\kern\dimen@#1}%
    \fi
  }%
  \macc@depth\@ne
  \let\math@bgroup\@empty \let\math@egroup\macc@set@skewchar
  \mathsurround\z@ \frozen@everymath{\mathgroup\macc@group\relax}%
  \macc@set@skewchar\relax
  \let\mathaccentV\macc@nested@a
  \if#31
    \macc@nested@a\relax111{#1}%
  \else
    \def\gobble@till@marker##1\endmarker{}%
    \futurelet\first@char\gobble@till@marker#1\endmarker
    \ifcat\noexpand\first@char A\else
      \def\first@char{}%
    \fi
    \macc@nested@a\relax111{\first@char}%
  \fi
  \endgroup
}
\makeatother

\def\bH{{\bf H}}

\def\bv{{\bf v}}

\def\bw{{\bf w}}

\def\bv{{\bf v}}

\def\ba{{\bf a}}

\def\be{{\bf e}}

\def\tt{{\mathrm{t}}}

\def\calB{{\mathcal{B}}}

\def\calL{{\mathcal{L}}}

\def\calV{{\mathcal{V}}}
\def\calW{{\mathcal{W}}}
\def\calU{{\mathcal{U}}}
\def\bzero{{\mathbf{0}}}
\def\HH{\mathsf{H}}
\def\Mb{M_{\text{BS}}}
\def\Mu{M_{\text{UE}}}
\def\Nb{N_{\text{BS}}}

\def\Nu{N_{\text{UE}}}
\def\Thetab{\theta_{\text{BS}}}
\def\Thetau{\theta_{\text{UE}}}
\def\Gb{G_{\text{BS}}}
\def\Gu{G_{\text{UE}}}

\def\b1{{\bf 1}}

\renewcommand{\IEEEQED}{\IEEEQEDopen}

\def\ES{\text{ES}}
\def\IS{\text{IS}}
\def\RB{\text{RB}}

\begin{document}
\graphicspath{{Components/Figures/}}

\title{Fast and Reliable Initial Access with Random Beamforming for mmWave Networks}

\author{Yanpeng Yang,~\IEEEmembership{Studnet Member,~IEEE,}
        Hossein S. Ghadikolaei,~\IEEEmembership{Member,~IEEE,}\\ 
        Carlo Fischione,~\IEEEmembership{Member,~IEEE,}
        Marina Petrova,~\IEEEmembership{Member,~IEEE,}
        and~Ki~Won~Sung,~\IEEEmembership{Member,~IEEE}
\thanks{Yanpeng Yang, Hossein S. Ghadikolaei, Carlo Fischione, Marina Petrova, and Ki Won Sung are with the School of Electrical Engineering and Computer Science, KTH Royal Institute of Technology, Stockholm, Sweden (e-mail: \{yanpeng, hshokri, carlofi, petrovam, sungkw\}@kth.se).}
}

\maketitle

\begin{abstract}
Millimeter-wave (mmWave) communications rely on directional transmissions to overcome severe path loss. Nevertheless, the use of narrow beams complicates the initial access procedure and increase the latency as the transmitter and receiver beams should be aligned for a proper link establishment. In this paper, we investigate the feasibility of random beamforming for the cell-search phase of initial access. We develop a stochastic geometry framework to analyze the performance in terms of detection failure probability and expected latency of initial access as well as total data transmission. Meanwhile, we compare our scheme with the widely used exhaustive search and iterative search schemes, in both control plane and data plane. Our numerical results show that, compared to the other two schemes, random beamforming can substantially reduce the latency of initial access with comparable failure probability in dense networks. We show that the gain of the random beamforming is more prominent in light traffics and low-latency services. Our work demonstrates that developing complex cell-discovery algorithms may be unnecessary in dense mmWave networks and thus shed new lights on mmWave network design.
\end{abstract}


\section{Introduction}
Millimeter-wave (mmWave) technology is one of the essential components of future wireless networks to support extremely high data rate services~\cite{Rappaport2013, pi2011introduction, Rangan2014Millimeter}. The mmWave frequency bands provides orders of magnitude more spectrum than the current congested bands below 6 GHz. It can offer much higher data rates and create a fertile ground for developing various new products and services \cite{Ghosh2014}. However, mmWave communications are subject to high path-loss, noise power and penetration loss \cite{Marcus2005}. 
To address the challenges, mmWave systems rely on directional transmissions using large antenna arrays both at the transmitter and at the receiver \cite{Roh2014}. 
Such directional transmission, albeit reduces the interference footprint and simplifies the scheduling task, complicates the initial synchronization and cell-search procedure which is a prerequisite to establish any connection in all cellular systems \cite{Shokri2015Transitional, Jeong2015Random}. The cell-search in initial access (IA) procedure of conventional cellular networks, e.g., LTE \cite{Dahlman2011}, is performed using omnidirectional antennas in the low frequency bands. However this is not applicable for mmWave communications due to the severe path-loss and the resulting mismatch in control and data-plane ranges \cite{Shokri-Ghadikolaei2015}. Consequently, it is essential to develop directional cell-search and IA.

\subsection{Related Work and Motivation}
The need for the design of new initial cell-search phase in mmWave communication has brought great research interest in this field recently. In IEEE 802.11ad standard, a coarse-grained sector matching is followed by a second beam training stage that provides a further refinement of the beamforming vectors \cite{6774849, Nitsche2014}. The authors in \cite{Desai2014} proposed a similar hierarchical design with multi-resolution codebook based on ideas from compressive sensing. Reference \cite{Giordani2018} provided a framework to evaluate the performance of mmWave IA using 3GPP new radio (NR) scenario configurations.
In \cite{Barati2016}, the authors analyzed various design options for IA given different scanning and signaling procedures. Specifically, the synchronization and cell-search consists of a sending a series of directional pilots to enable a joint time-frequency-spatial synchronization to occur jointly. \cite{Abbas2016} and \cite{Capone2015} investigated initial cell-search based on context information which uses external localization service to get positioning information. In \cite{Giordani2016}, the performance of these schemes are summarized in terms of cell detection failure probability and latency. In contrast to the aforementioned link-level studies, \cite{Li2017a} and \cite{Li2017b} provide a system-level analysis of IA protocols in terms of cell-search latency under different user equipment (UE) status. 

One of the fundamental drawbacks of almost all the algorithms above is that both base station (BS) and UE should sequentially scan the whole angular space (using a single-resolution or multi-resolution codebook) to find the \emph{best} link quality for the following data transmission. As the networks are becoming denser progressively, the potential data rate of directional mmWave communications is rather high even when serving by a BS other than the \emph{best} choice. Therefore, finding an \emph{acceptable} link especially in a dense network may be much more preferable, as it may substantially reduce the data-plane establishment latency while satisfying the rate demands of the UEs. Fast IA becomes even more important for machine type communications comprising many wake-up radios in which the synchronization may take longer than the actual data transmission, leading to a poor latency performance \cite{Jiang2018}. 
Random beamforming is an alternative scheme for directional cell-search in which the BSs (and UEs) focus their antenna patterns toward a randomly picked direction.
References\cite{Shokri-Ghadikolaei2015, Abu-Shaban2016, Barati2015directionall} have shown the feasibility of applying random beamforming for initial cell search of mmWave networks and its potential in providing low-latency IA. In particular, \cite{Shokri-Ghadikolaei2015} analyzed the delay statistics of initial access and showed promising results for the performance of random beamforming. In reference \cite{Abu-Shaban2016}, the authors drew a similar conclusion by analyzing the Cram\'{e}r-Rao lower bound for estimating directions of arrival and departure (essentially spatial synchronization). Reference \cite{Barati2015directionall} explored the boundaries of the SNR region where the synchronization signal is detectable. However, the studies are limited to either a deterministic channel model or single BS scenario. Our preliminary results \cite{yang2018} verified the benefits of random beamforming for the cell-search in dense mmWave networks under ideal assumptions for propagation environment (only line-of-sight (LOS) communications) and antenna models (only mainlobes).


\subsection{Contributions}
In this paper, we provide a system-level framework to analyze the performance of IA based on random beamforming in a multi-cell mmWave network. We substantially extended our initial results \cite{yang2018} by considering 3GPP NR framework, non-line-of-sight (NLOS) communications, non-zero antenna sidelobes, data plane performance, and extended numerical comparisons to alternative approaches. The main contributions of this paper are:
\begin{itemize}
\item \textbf{An analytical framework for IA performance under random beamforming:}
Leveraging the tools from stochastic geometry, we derive the exact expression of the detection failure probability and expected latency for initial access. Different from the previous works, we carry out a system-level analysis incorporating both sidelobe effect and NLOS paths. The analysis is validated by extensive Monto Carlo simulations.
\item \textbf{A detailed evaluation of the random beamforming for IA:}
We investigate the effect of BS density, environmental blockage and antenna beamwidth on IA performance. Meanwhile, we characterize the tradeoff between failure probability and expected latency. Through this, for any BS density, we find an optimal beamwidth that minimizes the expected latency subject to a detection failure constraint.
\item \textbf{A comparison to the widely-used exhaustive search and iterative search schemes:}
We show the superior IA latency performance under the random beamforming compared to the exhaustive search, especially in dense mmWave networks. Furthermore, the control plane and data plane overall latency is investigated. The proposed scheme outperforms the existing ones and the performance gain becomes more prominent with lighter traffics or shorter packet sizes. 
\end{itemize}

\subsection{Paper Organization}
The rest of the paper is organized as follows. In Section~\ref{sec: system-model}, we describe the system model. In Section~\ref{sec: Framework}, we present our initial access framework based on random beamforming. The analysis of detection failure probability and latency are presented in Section~\ref{sec: Analysis}. Simulation results and comparison to other schemes are provided in Section~\ref{sec: Numerical-results}. Section~\ref{sec: conclusions} concludes the paper.

\section{System Model}\label{sec: system-model}
In this section, we present the network, channel, and antenna models for evaluating the performance in this paper. Table 1 summarizes the main notations used throughout the paper.
\subsection{Network Model}
We consider a large-scale downlink mmWave cellular network where the BSs are distributed according to a two-dimensional Poisson point processes (PPP) $\Phi \triangleq \{x_i\}$ with density $\lambda$.  In \cite{Blaszczyszyn2015}, it has been shown that the PPP assumption can be viewed as incorporating BS locations and shadowing with sufficiently large variance. Therefore, we ignore the effect of shadowing as \cite{Park2016, bai2014coverage} in this work. The UEs follow another independent PPP, from which the typical UE, located at the origin, is our focus according to the Slivnyak's theorem~\cite{Haenggi2012}. In this paper, we focus on outdoor BSs and UEs by assuming the independence of the outdoor and indoor devices and invoking the thinning theorem \cite{Haenggi2012}.

\begin{table}
\caption{Notations and Simulation Parameters.}
\centering
\begin{tabular}{>{\centering}m{3cm} >{\centering}m{5cm} m{4cm}<{\centering}}
\hline \hline
Notation & Description & Simulation Value \\
\hline
$\Phi$ & BS PPP & \\
$\lambda$ & BS density & $\lambda = 10^{-5} \sim 10^{-3} /\text{m}^2$\\
$p_{\text{BS}}$ & BS transmit power & 30~dBm\\
$f_c$ & Operating frequency  & 28~GHz \\
$\Delta^c$ & Control plane bandwidth & 28.8~MHz\\
$\Delta^d$ & Data plane bandwidth & 100~MHz\\
NF & Noise figure & 7~dB\\
W & Thermal noise & -174~dBm/Hz + $10 \log_{10}(\Delta)$ \\
$\alpha_L$ & LOS path loss exponent  & 2.5\\
$\alpha_N$ & NLOS path loss exponent  & 4\\
$\beta$  & Blockage exponent   & 0.02\\
T & SINR threshold & 0 dB\\
$\Nb$  & No. of beamforming directions at BS & 12\\
$\Nu$  & No. of beamforming directions at UE & 4\\
$R_T$   & Achievable rate in data plane &  \\
$N_{ss}$ & No. of SS blocks in SS burst & 16/32/64\\
$T_{cs}$ & Cell-search (SS burst) duration  & 1.25/2.5/5~ms\\
$T_{ra}$ & Random access duration & 1.25~ms \\
$T_f$ & Frame duration  & 20~ms\\
$L$   & Transmission packet size & $10^3 \sim 10^9$ bits\\\hline \hline
\end{tabular}
\label{tab:para}
\end{table}

\subsection{Channel Model}
To the typical UE, each BS, independently from others, is characterized by either LOS or NLOS propagation. Define the LOS probability function $p_{\text{LOS}}(r)$ as the probability that a link of distance $r$ is in the LOS condition. We apply a stochastic exponential blockage model for the function where the obstacles are modeled by rectangle Boolean objects \cite{TBai2014Blockage}. In that case,  $p_{\text{LOS}(r)} = \exp(-\beta r)$ where $\beta$ is a parameter determined by the density and the average size of the obstacles, and $\beta^{-1}$ represents the average length of a LOS link. For the tractability of analysis, we further assume independent LOS events among different links \cite{TBai2014Blockage, TBai2014Coverage} and among different time slots~\cite{Li2017a}. 

Given LOS probability $p_{\text{LOS}}(r)$, the path loss for a link with distance $r$ is given by:

\begin{align}
\label{equ:pathloss}
\ell(\lVert x_i \rVert)=\left\{
        \begin{array}{ll}
          \(C(\lVert x_i \rVert)\)^{\alpha_L}, & \text{if LOS}; \\
          \(C(\lVert x_i \rVert)\)^{\alpha_N}, & \text{if NLOS (blocked)},
        \end{array}
      \right.
\end{align}
where $C(\lVert x_i \rVert) \triangleq c/4 \pi \lVert x_i \rVert f_c$, $c$ is the light speed, $f_c$ is the operating frequency, $\alpha_L$ and $\alpha_N$ represent the path loss exponent for LOS and NLOS links, and $\Vert x_i \Vert$ represents the Euclidean distance between $x_i \in \Phi$ and the origin $o$. To ignore the possibility of communications in the NLOS conditions, we can set $\alpha_N = \infty$ and $\alpha_L = \alpha$ for simplicity.

We assume that BSs and UEs are equipped with electronically-steered antenna arrays of $\Mb$ and $\Mu$ antennas respectively. Since mmWave channels are expected to have a limited number of scatterers \cite{Rappaport2013a}, we employ a geometric channel model with a single path between typical UE and each BS for better analytical tractability. The single-path assumption was implicitly adopted and verified in \cite{bai2014coverage, Alkhateeb2017}. The channel matrix between BS $b$ and UE $u$ is given by


\begin{equation}\label{eq: channel-matrix}
\bH_{ub} = \sqrt{\ell(\lVert x_i \rVert)} \, h_{ub} \, \ba\left(\Mu,\theta_{ub}\right) \ba^\HH \left(\Mb,\phi_{ub}\right) \:,
\end{equation}
where $h_{ub}$ represent the small-scale fading between BS $x_i$ and the typical UE. We assume $h_i$ follows a unit-mean Rayleigh distribution. Compared to more realistic models for LOS paths such as Nakagami fading, Rayleigh fading provides very similar design insights while it leads to more tractable results~\cite{Li2017a}. $\theta_{ub}\in [0, 2 \pi)$ and $\phi_{ub}\in [0, 2 \pi)$ are the angle of arrival (AoA) and the angle of departure (AoD) at UE $u$ and BS $b$ respectively, $(\cdot)^\HH$ is the conjugate transpose operator. Finally $\ba(k,\theta) \in \mathds{C}^{k}$ is the unit-norm vector response function of the transmitter's and receiver's antenna arrays to the AoAs and AoDs, given in (\ref{eq: ULA-antenna-response}).

\subsection{Antenna Model}
We consider analog beamforming for initial cell-search because digital or hybrid beamforming does not suit due to the existence of many antenna elements and lack of prior channel knowledge, translated into the need for costly pilot transmission schemes. Two antenna models are applied in this work. For analytical simplicity, we first model the actual antenna patterns by a sectorized beam pattern (SBP) as in~\cite{Shokri-Ghadikolaei2015}. We also consider the uniform linear array (ULA) antenna model in the numerical evaluations due to two reasons: 1) verifying the analytical insights and performance trends, obtained by the SBP model, and 2) obtaining the SINR in data plane with beam refinement, as shown in Section~\ref{sec:dataplane}. 

\subsubsection{Sectorized Beam Pattern}
 We consider half-power beamwidths of $\Thetab$ and $\Thetau$ at the BSs and UEs, respectively, with the corresponding antenna gains $\Gb$ and $\Gu$. In an ideal sectorized antenna pattern, the antenna gain $G_{\mathrm x}$, ${\mathrm x} \in \{{\text{BS}}, {\text{UE}}\}$, as a function of beamwidth $\theta_{\mathrm x}$ is a constant in the main lobe and a smaller constant in the side lobe, given by
\begin{align}\label{equ:antenna}
G_{\mathrm x}(\theta_{\mathrm x})=\left\{
        \begin{array}{ll}
          \frac{2\pi-(2\pi-\theta_{\mathrm x})\epsilon}{\theta_{\mathrm x}}, & \text{in the main lobe,} \\
          \epsilon, & \text{in the side lobe,} \\
        \end{array}
      \right.
\end{align}
where typically $\epsilon \ll 1$. For a given $\Thetab$ and $\Thetau$, which are a non-increasing function of the number of antenna elements, BSs and UEs sweep the entire angular space by $\Nb = \lceil2\pi/\Thetab\rceil$ and $\Nu = \lceil2\pi/\Thetau\rceil$ beamforming vectors, respectively. Without loss of generality of the main conclusions, we assume that $2\pi/\Thetab$ and $2\pi/\Thetau$ are integers and drop $\lceil\cdot\rceil$ operator. It is worth noting that we neglect the sidelobe gain at the UE side as of \cite{Alkhateeb2017} for mathematical tractability. 

\subsubsection{Uniform Linear Array}
 The array response vector can be expressed as
\begin{equation}\label{eq: ULA-antenna-response}
\ba\left(K,\theta\right) =
\frac{1}{\sqrt{k}}{
\begin{bmatrix}
1& e^{j \pi \sin(\theta)} & \ldots & e^{j (k - 1) \pi \sin(\theta)}
\end{bmatrix}^\HH} \:.
\end{equation}
The parameters of the channel model depend both on the carrier frequency and on being in LOS or NLOS conditions and are given in~\cite[Table~I]{Akdeniz2014MillimeterWave}.

To design the beamforming vectors (precoding at the BSs and combining at the UEs), we define  $\mathbf{f}(k,K,\theta)$ as
\begin{equation}\label{eq: P1-1}
\mathbf{f}(k,K,\theta) := \frac{
\begin{bmatrix}
1& e^{j \pi \sin(\theta)} & \ldots & e^{j (k - 1) \pi \sin(\theta)}& \bzero_{1\times(K - k)}
\end{bmatrix}^\HH}{\sqrt{k}}  \:,
\end{equation}
for integers $k$ and $K$ such that $0 < k \leq K$. $\bzero_x$ is an all zero vector of size $x$. Let $\bv_b^c$ and $\bw_u^c$ be the precoding vector of BS $b\in\calB$ and the combining vector of UE $u\in\calU$ in mini-slot $c$ of the cell-search phase. We define
\begin{subequations}\label{eq: precoder-combiner}
\begin{align}
\label{eq: precoder}
\bv_b^{c} & = \mathbf{f}(k_b^c,\Nb,\phi_b^c) \:, \\
\label{eq: combiner}
\bw_u^{c} & = \mathbf{f}(k_u^c,\Nu,\theta_u^c) \:.
\end{align}
\end{subequations}
The BSs and UEs can control the antenna boresight by changing $\phi_b^c$ and $\theta_u^c$ and control the antenna beamwidth by changing $k_b^c$ and $k_u^c$. At each BS $b$, we keep a local codebook $\calV_b^c$ that contains $\Mb$ precoding vectors. Each vector $\bv_b^c \in \calV_b^c$ is of the form~\eqref{eq: precoder} such that the codebook collectively spans all angular space. The cardinality of the codebook is based on the half-power beamwidth and determines the antenna gain and sidelobe interference caused by every beam. 


\subsection{3GPP NR Frame Model}
The 3GPP technical specification for NR introduces the concept of synchronization signal (SS) block and burst. An SS block spans four orthogonal frequency division multiplexing (OFDM) symbols in time and 240 subcarriers in frequency. See~\cite{Giordani2018} and references therein. Each SS block is mapped to a certain angular direction. An SS burst is a group of several SS blocks, and the interval between consecutive SS bursts $T_{\mathrm{SS}}$ can take $\{5,10,20,40,80,160\}$~ms. Higher values correspond to lower synchronization overhead. Within one frame of NR, there could be several pilots, called channel-state information reference signal (CSI-RS), to enable optimal beamforming design for the data-plane.

\begin{figure}[t]
\centering
\includegraphics[width=0.85\columnwidth]{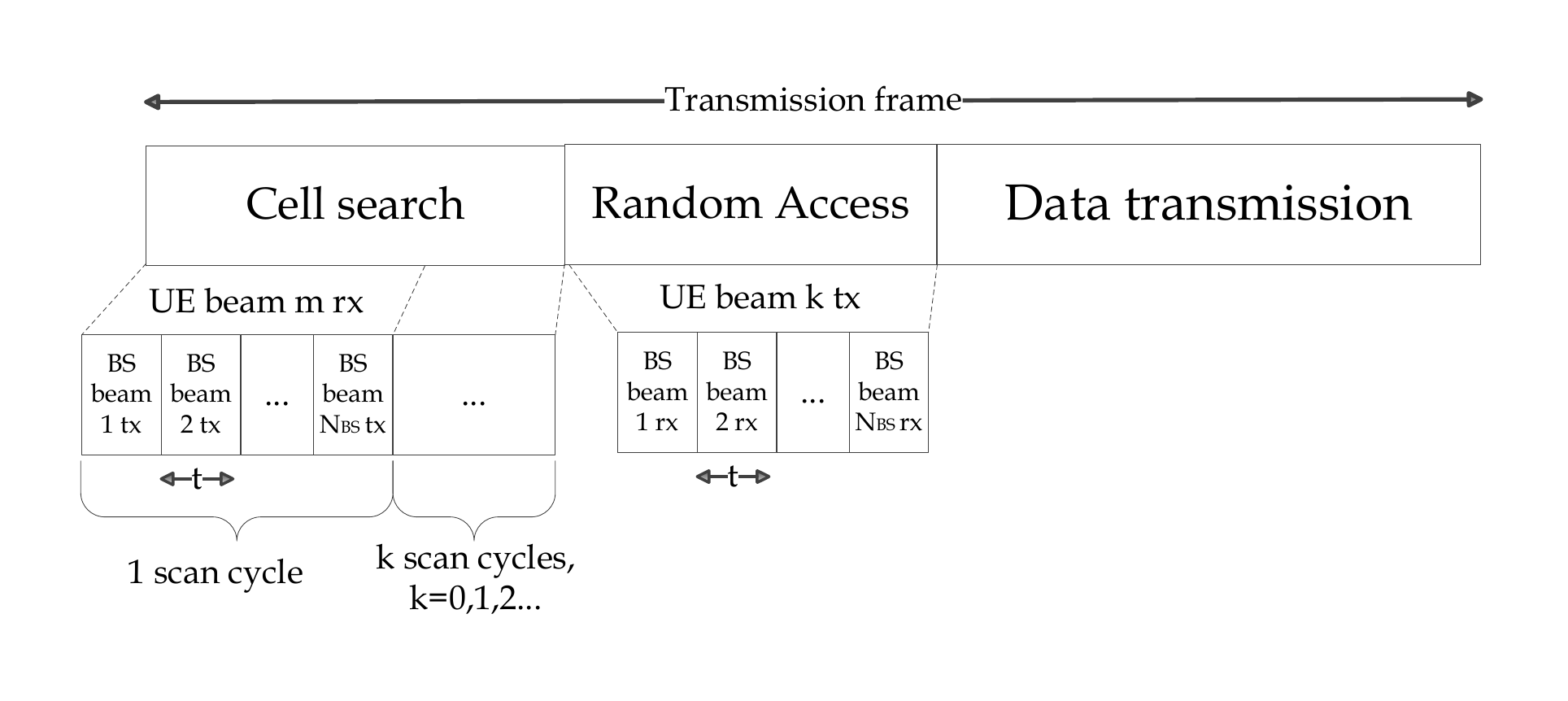}

\caption{Illustration of transmission frame with random beamforming.}
\label{fig:illu}
\end{figure}

\section{IA Framework under Random Beamforming and Performance Metrics}
\label{sec: Framework}
In this section, we introduce the initial access protocols and performance metrics. We assume the system time is divided into two phases within each coherence interval: 1) an initial access period comprising a cell search phase and a random access phase, 2) a data transmission period with beam alignment. The whole transmission frame is illustrated in Fig.~\ref{fig:illu}.

\subsection{Cell Search Phase}

 The cell-search period takes several mini-slots. Every BS independently and uniformly at random picks a direction out of $\Nb$ in each mini-slot. We define a scan cycle as the period within which every BS sends cell-search pilots to $\Nb$ directions, see Fig.~\ref{fig:illu}. In each scan cycle, the UE antenna points to a random direction out of $\Nu$, and the BS covers all non-overlapping $\Nb$ directions. Different from the exhaustive search and iterative search in which the BS and UE need to cover all the $\Nb \Nu$ possible directions \cite{Li2017a}, the cell-search period of random beamforming can be dynamically adjusted. Once the UE received a pilot signal that meets a predefined SINR threshold, it is associated to the corresponding BS. Note that it may not be the final association of that UE, but once the UE is registered to the network, it can establish data plane, and the reassociation phase (to the best BS) could be executed smoothly without service interruption~\cite{Shokri-Ghadikolaei2015}.

\subsection{Random Access Phase}
Given a successful cell search phase, the UE acquires the direction where it receives the strongest signal. In the following random access phase, the UE initiates the connection to its desired serving BS by transmitting random access preambles to that direction. The BS will scan for the presence of the random access preamble and will also learn the BF direction at the BS side. If cell search fails, the UE will skip the random access phase and repeat the cell search in the next frame.
In real systems, the UE picks the random access preamble from a certain number of orthogonal preamble sequences. The success of random access phase depends on: 1) no random access preamble collision for multiple UEs transmitting to the same BS; and 2) the SINR of the random access preamble signal exceeding a threshold. Since the main focus of this paper is random beamforming based cell search phase, the impact of random access performance is left in our future work. Therefore, we make the following assumptions:
\begin{assumption}
\label{assmp:collision}
    There is no random access preamble collision for all UEs, i.e. the BS can detect all the preambles.
\end{assumption}
\begin{assumption}
\label{asump:sinr}
    The random access phase and cell search phase share the same SINR threshold.
\end{assumption}    
Assumption~\ref{assmp:collision} holds in general, as the probability that multiple UEs pick the same random access preamble to access the same BS on the same spatial channel is very small, thanks to the low interference footprint of mmWave networks. Moreover, since the cell search and random access occur within the same coherence interval, Assumption~\ref{asump:sinr} implies that SINR in random access exceeds the threshold as long as the cell search phase succeeds. In general, under Assumption~\ref{assmp:collision} and \ref{asump:sinr}, we have simultaneous cell search and random access success or failure.  

\subsection{Data Plane Analysis}
\label{sec:dataplane}
After the RA phase, the connection is established and the data transmission starts.
To achieve more accurate performance evaluation, our data plane analysis is based on the ULA model. Let $\bv_{ub}^{dt}$ and $\bw_{u}^{dt}$ be the precoding vector of BS $b$ when serving UE $u$ and the combining vector of UE $u$ in the data transmission phase of coherence interval $t$, respectively. 
We assume that after the initial access phase UE $u$ and its serving BS $b$ will exchange a series of directional pilots, thanks to the available directional information from the initial access phase, to establish the data-plane with the maximum link budget:
\begin{subequations}\label{eq: AnalogBeamforming}
\begin{alignat}{3}
\label{eq: analog-beamforming-objective}
&\underset{\bv_{ub}^{dt},\bw_{u}^{dt}}{\text{maximize}} \hspace{3mm} && \left| \left(\bw_u^{dt}\right)^{\HH} \, \bH_{ub}^t \, \bv_{ub}^{dt} \right|^2 \:,
\\
\label{eq: analog-beamforming-C1}
&{\text{subject to}} && \bv_{ub}^{dt} \in \calV_b^{d}\:,
\\
\label{eq: analog-beamforming-C2}
& &&
\bw_u^{dt} \in \calW_u^{d} \:,
\end{alignat}
\end{subequations}
where $\calV_b^{d}$ and $\calW_u^{d}$ are the sets of feasible precoding and combining vectors: unit-norm, identical modulus, and of the form~\eqref{eq: P1-1}.

Afterward, the SINR of UE $u$ is
\begin{equation}\label{eq: SINR_dt}
\text{SINR}_{ub}^{dt} = \frac{ \left|\left(\bw_u^{dt}\right)^{\HH} \, \bH_{ub}^t \, \bv_{ub}^{dt}\right|^2 }{\sum\limits_{i\in \calB \setminus \{b\}} \left|\left(\bw_u^{dt}\right)^{\HH} \, \bH_{ui}^t \, \bv_{ui}^{dt}\right|^2 + \Delta^d N_0 / p_{\text{BS}}^d} \:.
\end{equation}
and the achievable rate is $R_T = \Delta^d \log\left(1+\text{SINR}_{ub}^{dt}\right)$, where $p_{\text{BS}}^d$ and $\Delta^d$ are the BS transmit power and the signal bandwidth of the data transmission phase, respectively. Note that the use of analog beamforming for the initial access does not pose any limitations on the beamforming architecture for the data-transmission phase. In other words, we can still use hybrid or digital beamforming for the data-plane. In that case, the SINR expression and achievable rates would be slightly different, though the tradeoffs and design insights of this paper (which are mostly focused on the performance of the control-plane and IA) are still valid.

\subsection{Performance Metrics}
Denoting $p_{\text{BS}}$ as the BS transmit power and $W$ as the thermal noise, the signal-to-interference-plus-noise ratio (SINR) when the typical UE is receiving from BS $x_i$ is given by
\begin{equation}
\label{eq: sinr}
\text{SINR}_i =\frac{ G_{\text{BS}_i} |h_{i}|^2 \ell(\lVert x_i \rVert) S_{i}}{\sum\limits_{x_j \in \Phi \backslash x_i} G_{\text{BS}_j} |h_{j}|^2 \ell(\lVert x_j \rVert)  S_{j} + \sigma^2} \:,
\end{equation}
where $\sigma^2=W(p_{\text{BS}} \Gu)^{-1}$ is the normalized noise power \footnote{In the propositions and proofs, we denote $\sigma^2 = W(p_{\text{BS}} \Gu (c/4 \pi f_c)^{\alpha_L}))^{-1}$ when $\alpha_N = \infty$.} and $S_i$ is LOS condition indicator.

We say the typical UE successfully detects the cell if the strongest signal it receives in any mini-slot from one of the directions achieves a minimum SINR threshold $T$. Namely, the success event is $\mathbb{I}\{\max_{x_i \in \Phi} \text{SINR}_{i} \geq T \}$, where $\mathbb{I}\{\cdot\}$ is the indicator function. For any realization of the topology $\Phi$ and channel fading $h$, detection failure may happen due to two reasons: there is no BS inside the UE's main beam or the ones inside cannot meet the detection threshold $T$.

Now, we define three performance metrics, for the typical UE, evaluated throughout the paper.
\begin{defi}\label{def: failure}
The detection failure probability $P_f(N_c)$ is the probability that the UE is not detected by any BS within $N_c$ mini-slots. 
\end{defi}

\begin{defi}\label{def: iadelay}
Given a time-budget of $N_c$ mini-slots, the IA latency $D_I(N_c)$ of a UE is defined as the time period by which the UE successfully detects a cell-search pilot and can be registered with the corresponding BS.
\end{defi}

\begin{defi}\label{def: datadelay}
Given a time-budget of $N_c$ mini-slots, the total data transmission latency $D_T(N_c)$ is defined as the time period during which the UE successfully registers with a BS and completes a packet transmission.
\end{defi}


\section{Detection failure probability and Latency Analysis}\label{sec: Analysis}

In this section, we present our analytical results and some insights on the performance of IA using random beamforming. Proofs are provided in the Appendix.

\subsection{Detection Failure Probability}
To get some insights about the problem, we start from a simple model where only mainlobe gain and LOS links are considered, i.e. $\epsilon = 0$ and $\alpha_N = \infty$. These assumptions are widely used in mmWave network analysis due to simplicity and the resulting acceptable accuracy of the performance analysis \cite{Shokri-Ghadikolaei2015}. Nonetheless, we discuss the effect of NLOS path and sidelobe later in this section.
In the following proposition, we derive the detection failure probability of initial cell search for the typical user under the simple model.

\begin{prop}\label{prop: failure-prob-mainlobe}
The detection failure probability $P_f(N_c)$ of the typical UE when $\epsilon = 0$ and $\alpha_N = \infty$ is given by:
\begin{equation}\label{eq: Pf}
P_{f}= (1- P_s)^{N_c} \:,
\end{equation}
where $P_s$ is the successful detection probability in one mini-slot, given by
\begin{equation}
   P_s = \frac{2\pi}{\Nb \Nu} \lambda \int_{0}^{\infty} \hspace{-2mm} e^{-T r^{\alpha} \sigma^2} \exp\(-\frac{2\pi}{\Nb \Nu} \lambda_b \int_0^{\infty} \frac{T r^{\alpha}e^{-\beta v}}{v^{\alpha}+T r^{\alpha}} v\mathrm{d}v\)  e^{-\beta r}  r \mathrm{d}r \:. 
\end{equation}
\end{prop}

It is worth noting that the failure probability will not fall to zero as $N_c \to \infty$. Due to blockage, there is always a non-zero probability that all the BSs are \emph{invisible} (i.e., blocked) to the typical UE. In this case, the initial access cannot succeed when assuming $\alpha_N = \infty$. However, in all scenarios of practical interests, $N_c$ cannot take very high values due to the corresponding overhead and latency.

Next, we incorporate NLOS paths into the analysis, i.e. $\alpha_L < \alpha_N < \infty$. The NLOS terms will be added into both signal and interference parts. We start from characterizing the laplace transform of the interference, followed by the derivation of the detection failure probability in Proposition~\ref{prop: failure-prob-NLoS}.

\begin{lem}\label{lem: laplace-nlos}
The Laplace transform of the interference when $\epsilon = 0$ and $\alpha_L \leq \alpha_N < \infty$ is given by: 
\begin{align}\label{eq: laplace-nlos}
\mathcal{L}_{I}^N(s) = \exp\(-\frac{2\pi}{\Nb \Nu} \lambda \int_{0}^{\infty}(1-\frac{e^{-\beta v} v^{\alpha_L}}{v^{\alpha_L}+s k_1} - \frac{(1-e^{-\beta v}) v^{\alpha_N}}{v^{\alpha_N}+s k_2})v\mathrm{d}v\),
\end{align}
where $k_1 = \(\frac{c}{4 \pi f_c}\)^{\alpha_L}, k_2 = \(\frac{c}{4 \pi f_c}\)^{\alpha_N}$.
\end{lem}

\begin{prop}\label{prop: failure-prob-NLoS}
The detection failure probability $P_f^N(N_c)$ of the typical UE when $\epsilon = 0$ and $\alpha_L \leq \alpha_N < \infty$ is:
\begin{equation}\label{eq: Pfnlos}
P_f^{N}= (1- P_s^N)^{N_c} \:,
\end{equation}
where $P_s^N$ is the successful detection probability in one mini-slot, given by
\begin{align}
P_s^N &=\frac{2\pi}{\Nb \Nu} \lambda  \int_{0}^{\infty} \( \kappa_L + \kappa_N \) r \mathrm{d}r, \\
\kappa_L &= e^{-T C(r)^{-\alpha_L} \sigma^2} \calL_{I}(T C(r)^{-\alpha_L}) e^{-\beta r} \nonumber, \\
\kappa_N &= e^{-T C(r)^{-\alpha_N} \sigma^2} \calL_{I}(T C(r)^{-\alpha_N}) (1-e^{-\beta r}) \nonumber.
\end{align}
\end{prop}

Although Proposition~\ref{prop: failure-prob-NLoS} appears unwieldy, we may gain some insight by decomposing the terms therein. In Proposition~\ref{prop: failure-prob-NLoS}, $\kappa_L$ and $\kappa_N$ correspond to the LOS and NLOS contributions to the coverage, respectively. When $r$ is small, the LOS probability is higher, i.e. $e^{-\beta r} > 1 - e^{-\beta r}$. Meanwhile, the LOS signal strength also dominates that of NLOS, which makes overall $\kappa_L \gg \kappa_N$. Furthermore, as $r$ becomes larger, although the NLOS probability grows to 1, $\kappa_N$ is still very small, as the signal strength of a far-away NLOS BS is comparable or even smaller than the noise power. These insights indicate that NLOS signals have limited impact on the performance as \cite{Thornburg2016, TBai2014Coverage}, which is also validated in our numerical results in Section~\ref{sec: Numerical-results}.   

At last, the detection failure probability considering sidelobe gain $0 \leq \epsilon \leq 1$ is considered. When taking the sidelobe gain into account, we cannot treat each mini-slot independently as in Propositions~\ref{prop: failure-prob-mainlobe} and \ref{prop: failure-prob-NLoS} due to the correlation caused by sidelobes. In this situation, we consider the BSs pointing to the typical UE with mainlobe and sidelobe as two independent tiers of PPP distributed BSs as in \cite{Dhillon2012}. Thus, we divide the analsis into mainlobe and sidelobe failure. The accuracy is validated in Section~\ref{sec: Numerical-results}.

\begin{lem}\label{lem: mainlobe-success}
The probability successfully detected by the mainlobe in one mini-slot when $0 \leq \epsilon \leq 1$ and $\alpha_N = \infty$ is:
\begin{align}\label{eq: mainlobe-success}
P_{sm} = \Thetau \frac{\Thetab}{2\pi} \lambda \int_{0}^{\infty} \mathcal{L}_{I_{x_i}} (\frac{T r^{\alpha}}{G_{\text{BS}_i}}) e^{-T r^{\alpha} \sigma^2} e^{-\beta r} r\mathrm{d}r \:,
\end{align}
where $\mathcal{L}_{I_{x_i}}^m (s)$ is the Laplace transform of the interference given in (\ref{eq:laplace-mainlobe1}).
\end{lem}

\begin{lem}\label{lem: sidelobe-success}
The probability successfully detected by the sidelobe over $n < \Nb$ mini-slots when $0 \leq \epsilon \leq 1$ and $\alpha_N = \infty$ is:
\begin{equation}
    P_{ss} = \Thetau \frac{2\pi - \Thetab}{2\pi} \lambda \int_{0}^{\infty} Q_{x_i}^n(T)  e^{-\beta r} r\mathrm{d}r \:,
\end{equation}
where $Q_{x_i}^n(T)$ represents the probability of at least one successful detection during the $n$ mini-slots at a certain BS $x_i$, given in (\ref{eq:selection}).
\end{lem}

\begin{prop}\label{prop: detection-failure-prob-sidelobe}
The detection failure probability $P_f^S(N_c)$ of a typical UE when $0 < \epsilon < 1$ and $\alpha_N = \infty$ can be approximated as:
\begin{equation}\label{eq: Pfsidelobe}
P_f^{\text{S}} = (1- P_{ss})(1- P_{sm})^{N_c} \:,
\end{equation}
where $P_{sm}$ and $P_{ss}$ are the successful detection probabilities by the mainlobe within one mini-slot and by the sidelobe over $N_c$ mini-slots respectively, given in the Lemma~\ref{lem: mainlobe-success} and \ref{lem: sidelobe-success}.
\end{prop}

It is common in the literature to neglect the antenna sidelobes due to its large gap to the mainlobe and for analytical simplicity \cite{Li2017b}. However, sidelobes may play an important role when the BSs and obstacles are getting denser. In Proposition~\ref{prop: detection-failure-prob-sidelobe}, $P_{ss}$ grows with BS density, thereby contributing to successful detections.

Proposition \ref{prop: failure-prob-mainlobe} to \ref{prop: detection-failure-prob-sidelobe} give the expressions for the detection failure probability under three different scenarios. We observe that the failure probability depends on the result of each time slot $P_s$ and the total time budget $N_c$. From the proofs, $P_s$ is further characterized by the BS density, beamwidth and blockage. Among these parameters, increasing the BS density and time budget can reduce the failure probability by enlarging the search space. The effect of beamwidth and blockage is not straightforward. Narrowing the beamwidth enhances the beamforming gain but leads to fewer available BSs. Lighter blockage leaves more available LOS BSs, yet creates a stronger interference. More details will be discussed in Section~\ref{sec: Numerical-results}. 

\subsection{Delay Analysis}
Next, we derive the expected latency of the IA and data transmission. Recall notations $T_f$, $T_{\text{cs}}$, $T_{\text{ra}}$, and Definitions~\ref{def: iadelay} and \ref{def: datadelay}. We have the following proposition.  

\begin{prop}\label{prop: expected-latency}
The expected initial access latency $D_I(N_c)$ and total data transmission latency $D_T(N_c)$ of a typical UE are:
\begin{align}
\mathbb{E}[D_I(N_c)] &= \(\frac{1}{1-P_f(N_c)} - 1\)T_{f} + T_{\text{cs}} + T_{\text{ra}} \label{eq: ialatency},\\
\mathbb{E}[D_T(N_c)] &= \(\frac{1}{1-P_f(N_c)} - 1\)T_{f} + \left\lceil \frac{L}{R_T(T_f- T_{\text{cs}} - T_{\text{ra}})} \right\rceil (T_{\text{cs}} + T_{\text{ra}}) + \frac{L}{R_T} \label{eq: datalatency},\:
\end{align}
where $L$ is the packet size for transmission and $R_T$ is the achievable rate given in Section~\ref{sec:dataplane}.
\end{prop}


Considering the characterizations of the detection failure probability and IA latency, we aim to design the BS beamwidth ($\Thetab$ or equivalently $\Nb$) to minimize the IA latency $D_I(N_c)$ given a failure probability constraint $P_f^{\text{max}} \in [0,1]$. 
To formulate the problem, we set the cell-search time budget as $k$ scan cycle for all beamwidths, i.e. $N_c= k \Nb$ mini-slots. The optimal number of sectors for $k\in\mathbb{N}$ scan cycles is
\begin{subequations}\label{eq: optimization-problem}
\begin{alignat}{3}
\min_{\Nb}&  \hspace{3mm} && \mathbb{E}[D_I(k\Nb)]  \\
\text{s.t.} & && P_f(k\Nb) \leq P_f^{\text{max}}, \\
& && \Nb \in \mathbb{N} \:,
\end{alignat}
\end{subequations}
where $P_f^{\text{max}} $ and $k$ are inputs to this optimization problem. In \eqref{eq: ialatency}, $T_{\text{cs}}$ and $T_{\text{ra}}$ are linearly increasing functions of $\Nb$ while $\(\frac{1}{1-P_f(N_c)} - 1\)T_{f}$ is a decreasing function of $\Nb$ with a diminishing slope. Consequently the latency decreases with $\Nb$ up to some point and increases afterward. Therefore, we can always find the optimal solution of \eqref{eq: optimization-problem} by searching over rather small $\Nb$ values. The optimization problem~\eqref{eq: optimization-problem} can be utilized for system design in mmWave networks. With the knowledge of network deployment (BS density), we can set the configuration of BS antennas to meet the requirements of various applications with different reliability and/or latency constraints. An example of the solutions to the problem can be observed from the figures in the next section.

\section{Numerical Results}\label{sec: Numerical-results}
In this section, we present the numerical results of IA based on random beamforming. The simulation parameters are summarized in Table~\ref{tab:para}. In the figures, "\RB", "\IS", and "\ES" stand for random beamforming, iterative search and exhaustive search, respectively. In this paper, we focus on $N_c = \Nb$, i.e. the random cell-search lasts for 1 scan cycle. The effect of multiple cycles is present in our previous work \cite{yang2018}. We compare our scheme with exhaustive search and iterative search in a 3GPP NR framework. For a fair comparison, we set the number of SS blocks in one SS burst to 16, 32, and 64 for random beamforming, iterative search and exhaustive search so that every scheme can complete one scan cycle within one transmission frame. The beamwidth in the first stage of iterative search is set to $\theta_{\text{BS1}}=\ang{90}$.
As we mentioned in Introduction, there exists other cell-search algorithms either working on the link-level or designed for a single-cell scenario, which are not fairly comparable with our system-level multi-cell scenario and thus comparison with them is left for future work.

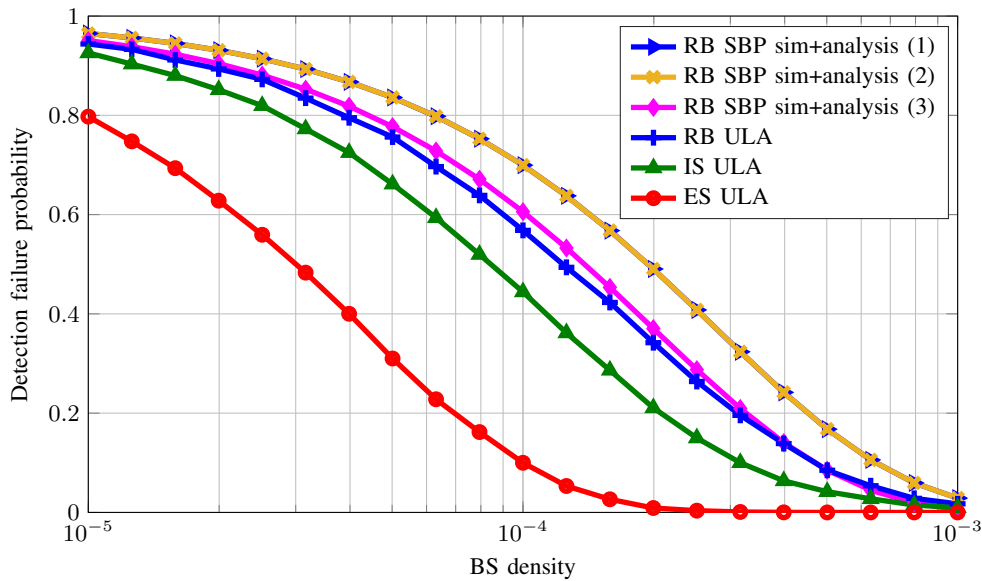
\begin{figure}[t]
\centering
\footnotesize{
%
%
\definecolor{mycolor1}{rgb}{0.92941,0.69412,0.12549}%
\definecolor{mycolor2}{rgb}{0.75000,0.00000,0.75000}%
\definecolor{mycolor3}{rgb}{1.00000,0.00000,1.00000}%
\begin{tikzpicture}

\begin{axis}[%
width=0.7\columnwidth,
height=0.4\columnwidth,
at={(0\columnwidth,0\columnwidth)},
scale only axis,
xmode=log,
xmin=1e-05,
xmax=0.001,
xminorticks=true,
xlabel={BS density},
ymin=0,
ymax=1,
ylabel={Detection failure probability},
axis background/.style={fill=white},
xlabel near ticks,
ylabel near ticks,
xmajorgrids,
xminorgrids,
ymajorgrids,
legend style={at={(0.99,0.98)},anchor=north east,legend cell align=left,align=left,draw=white!15!black,legend columns=1}
]
\addplot [color=blue, line width=1.0pt, forget plot]
  table[row sep=crcr]{%
1e-05	0.96506\\
1.25892541179417e-05	0.95661\\
1.58489319246111e-05	0.94604\\
1.99526231496888e-05	0.92985\\
2.51188643150958e-05	0.91426\\
3.16227766016838e-05	0.89481\\
3.98107170553497e-05	0.86668\\
5.01187233627272e-05	0.83417\\
6.30957344480193e-05	0.79506\\
7.94328234724282e-05	0.75506\\
0.0001	0.69798\\
0.000125892541179417	0.63686\\
0.000158489319246111	0.5713\\
0.000199526231496888	0.49005\\
0.000251188643150958	0.40683\\
0.000316227766016838	0.32001\\
0.000398107170553497	0.24079\\
0.000501187233627273	0.1659\\
0.000630957344480193	0.10679\\
0.000794328234724281	0.05879\\
0.001	0.02892\\
};
\addplot [color=blue, line width=2.0pt, mark=triangle, mark options={solid, rotate=270, blue}]
  table[row sep=crcr]{%
1e-05	0.96488156453067\\
1.25892541179417e-05	0.955991950534156\\
1.58489319246111e-05	0.944917337571697\\
1.99526231496888e-05	0.931158065972533\\
2.51188643150958e-05	0.91412163477994\\
3.16227766016838e-05	0.893118004501615\\
3.98107170553497e-05	0.867363106539633\\
5.01187233627272e-05	0.835996496331888\\
6.30957344480193e-05	0.798121416977844\\
7.94328234724282e-05	0.752877847610344\\
0.0001	0.699560374421911\\
0.000125892541179417	0.637790977965751\\
0.000158489319246111	0.567748481898457\\
0.000199526231496888	0.490437635577314\\
0.000251188643150958	0.407946458523175\\
0.000316227766016838	0.323593301579388\\
0.000398107170553497	0.241819089141096\\
0.000501187233627273	0.167674763001715\\
0.000630957344480193	0.105848582043229\\
0.000794328234724281	0.0594151064456971\\
0.001	0.0288043315576723\\
};
\addlegendentry{RB SBP sim+analysis (1)}

\addplot [color=mycolor1, forget plot]
  table[row sep=crcr]{%
1e-05	0.96522\\
1.25892541179417e-05	0.95653\\
1.58489319246111e-05	0.94449\\
1.99526231496888e-05	0.93145\\
2.51188643150958e-05	0.91586\\
3.16227766016838e-05	0.89427\\
3.98107170553497e-05	0.86621\\
5.01187233627272e-05	0.83474\\
6.30957344480193e-05	0.7956\\
7.94328234724282e-05	0.75322\\
0.0001	0.6968\\
0.000125892541179417	0.64049\\
0.000158489319246111	0.56934\\
0.000199526231496888	0.48991\\
0.000251188643150958	0.40589\\
0.000316227766016838	0.32095\\
0.000398107170553497	0.23999\\
0.000501187233627273	0.16866\\
0.000630957344480193	0.10518\\
0.000794328234724281	0.05745\\
0.001	0.02948\\
};
\addplot [color=mycolor1, line width=2.0pt, mark size=3.0pt, mark=x, mark options={solid, mycolor1}]
  table[row sep=crcr]{%
1e-05	0.96487997412697\\
1.25892541179417e-05	0.955989967259096\\
1.58489319246111e-05	0.944914870449413\\
1.99526231496888e-05	0.931155006446842\\
2.51188643150958e-05	0.914117855359296\\
3.16227766016838e-05	0.893113358643051\\
3.98107170553497e-05	0.867357430767743\\
5.01187233627272e-05	0.835989616022453\\
6.30957344480193e-05	0.798113157624305\\
7.94328234724282e-05	0.752868054248412\\
0.0001	0.699548946488126\\
0.000125892541179417	0.637777887219883\\
0.000158489319246111	0.567733857457266\\
0.000199526231496888	0.490421794892497\\
0.000251188643150958	0.407929954755609\\
0.000316227766016838	0.323576927721821\\
0.000398107170553497	0.241803813255471\\
0.000501187233627273	0.167661571824852\\
0.000630957344480193	0.105838245384741\\
0.000794328234724281	0.0594079360267192\\
0.001	0.0288000619139252\\
};
\addlegendentry{RB SBP sim+analysis (2)}

\addplot [color=mycolor2, forget plot]
  table[row sep=crcr]{%
1e-05	0.95062\\
1.25892541179417e-05	0.9385\\
1.58489319246111e-05	0.92299\\
1.99526231496888e-05	0.90392\\
2.51188643150958e-05	0.88172\\
3.16227766016838e-05	0.8536\\
3.98107170553497e-05	0.81789\\
5.01187233627272e-05	0.77745\\
6.30957344480193e-05	0.72823\\
7.94328234724282e-05	0.67017\\
0.0001	0.60567\\
0.000125892541179417	0.52977\\
0.000158489319246111	0.45236\\
0.000199526231496888	0.36788\\
0.000251188643150958	0.28675\\
0.000316227766016838	0.20837\\
0.000398107170553497	0.14222\\
0.000501187233627273	0.08638\\
0.000630957344480193	0.04712\\
0.000794328234724281	0.02235\\
0.001	0.00961\\
};
\addplot [color=mycolor3, line width=2.0pt, mark=diamond, mark options={solid, mycolor3}]
  table[row sep=crcr]{%
1e-05	0.950778590252769\\
1.25892541179417e-05	0.938446246109506\\
1.58489319246111e-05	0.923155768203573\\
1.99526231496888e-05	0.904271365346839\\
2.51188643150958e-05	0.881063131553263\\
3.16227766016838e-05	0.852717341490409\\
3.98107170553497e-05	0.818365107650646\\
5.01187233627272e-05	0.777138187647838\\
6.30957344480193e-05	0.728262654652784\\
7.94328234724282e-05	0.671200689414881\\
0.0001	0.605845579065448\\
0.000125892541179417	0.532761608121488\\
0.000158489319246111	0.453435210149306\\
0.000199526231496888	0.370466027989758\\
0.000251188643150958	0.287585108003243\\
0.000316227766016838	0.209367505112146\\
0.000398107170553497	0.140550968932013\\
0.000501187233627273	0.0850201642974136\\
0.000630957344480193	0.0447517182819575\\
0.000794328234724281	0.0192070555166395\\
0.001	0.0055885814813564\\
};
\addlegendentry{RB SBP sim+analysis (3)}

\addplot [color=blue, line width=2.0pt, mark size=3.0pt, mark=+, mark options={solid, blue}]
  table[row sep=crcr]{%
1e-05	0.94375\\
1.25892541179417e-05	0.9329\\
1.58489319246111e-05	0.91165\\
1.99526231496888e-05	0.8935\\
2.51188643150958e-05	0.8725\\
3.16227766016838e-05	0.8347\\
3.98107170553497e-05	0.79515\\
5.01187233627272e-05	0.7563\\
6.30957344480193e-05	0.69675\\
7.94328234724282e-05	0.63875\\
0.0001	0.56825\\
0.000125892541179417	0.4938\\
0.000158489319246111	0.423\\
0.000199526231496888	0.34195\\
0.000251188643150958	0.26375\\
0.000316227766016838	0.1956\\
0.000398107170553497	0.1387\\
0.000501187233627273	0.08495\\
0.000630957344480193	0.05375\\
0.000794328234724281	0.02835\\
0.001	0.01725\\
};
\addlegendentry{RB ULA}

\addplot [color=black!50!green, line width=2.0pt, mark=triangle, mark options={solid, black!50!green}]
  table[row sep=crcr]{%
1e-05	0.92575\\
1.25892541179417e-05	0.90325\\
1.58489319246111e-05	0.8801\\
1.99526231496888e-05	0.85155\\
2.51188643150958e-05	0.81945\\
3.16227766016838e-05	0.7726\\
3.98107170553497e-05	0.72485\\
5.01187233627272e-05	0.6614\\
6.30957344480193e-05	0.59365\\
7.94328234724282e-05	0.5196\\
0.0001	0.44425\\
0.000125892541179417	0.3616\\
0.000158489319246111	0.28635\\
0.000199526231496888	0.2106\\
0.000251188643150958	0.1499\\
0.000316227766016838	0.1001\\
0.000398107170553497	0.0636\\
0.000501187233627273	0.04175\\
0.000630957344480193	0.0275\\
0.000794328234724281	0.01495\\
0.001	0.0094\\
};
\addlegendentry{IS ULA}

\addplot [color=red, line width=2.0pt, mark=o, mark options={solid, red}]
  table[row sep=crcr]{%
1e-05	0.7974\\
1.25892541179417e-05	0.74755\\
1.58489319246111e-05	0.69355\\
1.99526231496888e-05	0.6281\\
2.51188643150958e-05	0.5594\\
3.16227766016838e-05	0.48305\\
3.98107170553497e-05	0.4002\\
5.01187233627272e-05	0.3101\\
6.30957344480193e-05	0.22795\\
7.94328234724282e-05	0.1619\\
0.0001	0.1\\
0.000125892541179417	0.05325\\
0.000158489319246111	0.02645\\
0.000199526231496888	0.0089\\
0.000251188643150958	0.00355\\
0.000316227766016838	0.00095\\
0.000398107170553497	0.00015\\
0.000501187233627273	0\\
0.000630957344480193	0\\
0.000794328234724281	0\\
0.001	0\\
};
\addlegendentry{ES ULA}

\end{axis}

\begin{axis}[%
width=0.7\columnwidth,
height=0.4\columnwidth,
at={(0\columnwidth,0\columnwidth)},
scale only axis,
xmin=0,
xmax=1,
ymin=0,
ymax=1,
axis line style={draw=none},
ticks=none,
axis x line*=bottom,
axis y line*=left,
legend style={legend cell align=left, align=left, draw=white!15!black}
]
\end{axis}
\end{tikzpicture}%
}
\caption{Density effect on failure probability. (1) refers $\epsilon=0, \alpha_N = \infty$, (2) refers $\epsilon=0, \alpha_L \leq \alpha_N < \infty$, and (3) refers $0 \leq \epsilon \leq 1, \alpha_N = \infty$.}
\label{fig:pf}
\end{figure}

Fig.~\ref{fig:pf} shows the cell-search performance against BS density $\lambda$. We observe a good matching between our theoretical analysis and the Monte Carlo simulations with the SBP model. Moreover, the difference is rather small between the curves using the SBP and ULA models. It is shown that the effect of NLOS paths on failure probability is negligible due to the overlapping of curves (1) and (2). Nevertheless, the sidelobe gain plays an important role which causes a notable decrease on detection failure probability. From Fig.~\ref{fig:pf}, the detection failure probability reduces with the BS density for all schemes. As BS density grows, the gap between random beamforming and other schemes diminishes rapidly. In dense regime where BS density approaches $10^{-3} / \text{m}^2$, all the probabilities converge to 0 gradually.  

\begin{figure}
\centering
\footnotesize{
%
%
\begin{tikzpicture}

\begin{axis}[%
width=0.7\columnwidth,
height=0.4\columnwidth,
at={(0\columnwidth,0\columnwidth)},
scale only axis,
xmode=log,
xmin=1e-05,
xmax=0.001,
xminorticks=true,
xlabel={BS density},
ymode=log,
ymin=1,
ymax=1000,
yminorticks=true,
ylabel={Expected IA latency [ms]},
axis background/.style={fill=white},
xlabel near ticks,
ylabel near ticks,
xmajorgrids,
xminorgrids,
ymajorgrids,
yminorgrids,
legend style={at={(0.02,0.05)},anchor=south west,legend cell align=left,align=left,draw=white!15!black,legend columns=-1}
]
\addplot [color=blue, line width=2.0pt, mark size=3.0pt, mark=+, mark options={solid, blue}]
  table[row sep=crcr]{%
1e-05	338.055555555555\\
1.25892541179417e-05	280.56259314456\\
1.58489319246111e-05	208.872382569326\\
1.99526231496888e-05	170.293427230047\\
2.51188643150958e-05	139.362745098039\\
3.16227766016838e-05	103.492135511192\\
3.98107170553497e-05	80.1324139614352\\
5.01187233627272e-05	64.5681165367255\\
6.30957344480193e-05	48.4521846661171\\
7.94328234724282e-05	37.863321799308\\
0.0001	28.8231036479444\\
0.000125892541179417	22.0100750691426\\
0.000158489319246111	17.1620450606586\\
0.000199526231496888	12.8928272927589\\
0.000251188643150958	9.66468590831918\\
0.000316227766016838	7.36325211337643\\
0.000398107170553497	5.72071287588529\\
0.000501187233627273	4.35672914048413\\
0.000630957344480193	3.63606340819022\\
0.000794328234724281	3.08354345700612\\
0.001	2.85105571101501\\
};
\addlegendentry{RB}

\addplot [color=black!50!green, line width=2.0pt, mark=triangle, mark options={solid, black!50!green}]
  table[row sep=crcr]{%
1e-05	253.110269360269\\
1.25892541179417e-05	190.46834625323\\
1.58489319246111e-05	150.555671392827\\
1.99526231496888e-05	118.475496800269\\
2.51188643150958e-05	94.5226391581279\\
3.16227766016838e-05	71.7007475813544\\
3.98107170553497e-05	56.4376249318554\\
5.01187233627272e-05	42.8167454223272\\
6.30957344480193e-05	32.9686538698167\\
7.94328234724282e-05	25.381973355537\\
0.0001	19.737404408457\\
0.000125892541179417	15.078320802005\\
0.000158489319246111	11.7749421985567\\
0.000199526231496888	9.08569799847986\\
0.000251188643150958	7.27664392424421\\
0.000316227766016838	5.9746916324036\\
0.000398107170553497	5.10839384878257\\
0.000501187233627273	4.62138012001044\\
0.000630957344480193	4.31555269922879\\
0.000794328234724281	4.05353789147759\\
0.001	3.93978396931153\\
};
\addlegendentry{IS}

\addplot [color=red, line width=2.0pt, mark=o, mark options={solid, red}]
  table[row sep=crcr]{%
1e-05	84.9666831194472\\
1.25892541179417e-05	65.4736086353734\\
1.58489319246111e-05	51.5135013868494\\
1.99526231496888e-05	40.0278972842162\\
2.51188643150958e-05	31.6426463912846\\
3.16227766016838e-05	24.9384611664571\\
3.98107170553497e-05	19.5944481493831\\
5.01187233627272e-05	15.239708653428\\
6.30957344480193e-05	12.1550579625672\\
7.94328234724282e-05	10.1135007755638\\
0.0001	8.47222222222222\\
0.000125892541179417	7.37490097702667\\
0.000158489319246111	6.79337219454573\\
0.000199526231496888	6.42959842599132\\
0.000251188643150958	6.32125294796528\\
0.000316227766016838	6.26901806716381\\
0.000398107170553497	6.25300045006751\\
0.000501187233627273	6.25\\
0.000630957344480193	6.25\\
0.000794328234724281	6.25\\
0.001	6.25\\
};
\addlegendentry{ES}

\end{axis}

\begin{axis}[%
width=0.7\columnwidth,
height=0.4\columnwidth,
at={(0\columnwidth,0\columnwidth)},
scale only axis,
xmin=0,
xmax=1,
ymin=0,
ymax=1,
axis line style={draw=none},
ticks=none,
axis x line*=bottom,
axis y line*=left,
legend style={legend cell align=left, align=left, draw=white!15!black}
]
\end{axis}
\end{tikzpicture}%
}
\caption{BS density effect on IA delay (ULA model).}
\label{fig:iadelay}
\end{figure}
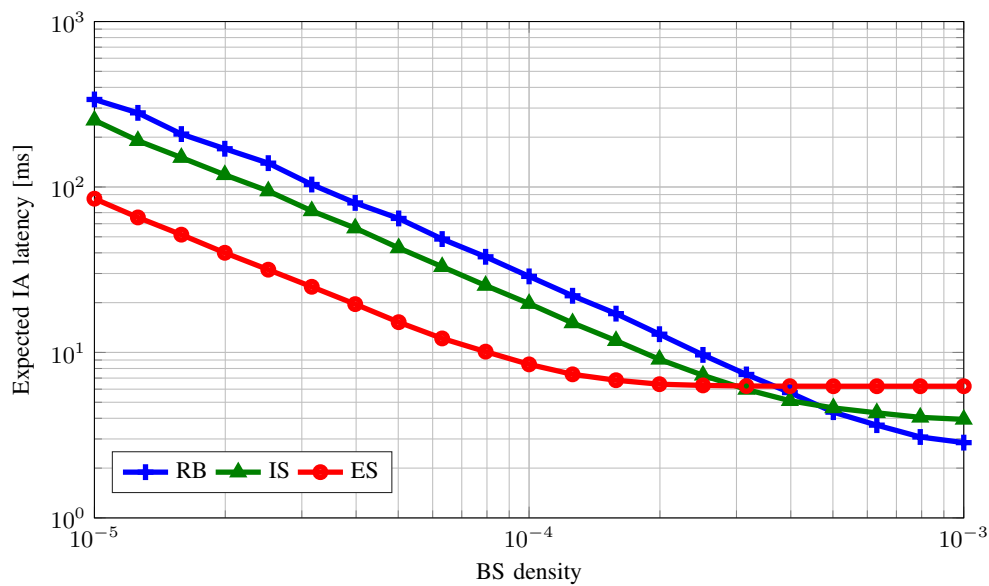

Fig.~\ref{fig:iadelay} compares the expected IA latency of three schemes over different BS density regimes. In the low density regime, the IA latency of exhaustive search is the shortest even though its cell-search period is the longest. This is because it consumes the least number of overhead frames before transmission due to low failure probability. However, the IA latency tends to converge to a constant as the failure probability converges to 0. On the other hand, the IA latency of random beamforming keeps reducing as more BSs are deployed. When the BS density exceeds $200/\text{km}^2$, the IA latency of the random beamforming becomes lower than that of the exhaustive and iterative search schemes. The main reason is the availability of more candidate BSs to register the UE. Note that the best possible values for the failure probability and normalized expected latency (number of mini-slot) are 0 and 1, respectively. By employing random beamforming, we can get close to those values in dense mmWave networks, alleviating the need for optimal (and perhaps complicated) cell-search procedures.

\begin{figure}
\centering
\footnotesize{
%
%
\definecolor{mycolor1}{rgb}{0.75000,0.00000,0.75000}%
\definecolor{mycolor2}{rgb}{1.00000,0.00000,1.00000}%
\begin{tikzpicture}

\begin{axis}[%
width=0.7\columnwidth,
height=0.4\columnwidth,
at={(0\columnwidth,0\columnwidth)},
scale only axis,
xticklabel style={
  /pgf/number format/precision=3,
  /pgf/number format/fixed},
xmin=0,
xmax=0.1,
xlabel={$\text{Blockage exponent }\beta$},
ymin=0,
ymax=1,
ylabel={Detection failure probability},
axis background/.style={fill=white},
xlabel near ticks,
ylabel near ticks,
xmajorgrids,
ymajorgrids,
legend style={at={(0.98,0.02)},anchor=south east,legend cell align=left,align=left,draw=white!15!black,legend columns=1}
]
\addplot [color=blue, line width=2.0pt]
  table[row sep=crcr]{%
0	0.0148\\
0.005	0.0996\\
0.01	0.3117\\
0.015	0.486\\
0.02	0.6115\\
0.025	0.6761\\
0.03	0.7337\\
0.035	0.7651\\
0.04	0.7959\\
0.045	0.8199\\
0.05	0.8317\\
0.055	0.8533\\
0.06	0.8639\\
0.065	0.8747\\
0.07	0.8908\\
0.075	0.893\\
0.08	0.9094\\
0.085	0.9144\\
0.09	0.9232\\
0.095	0.93\\
0.1	0.9337\\
};
\addlegendentry{$\text{N}_{\text{BS}}\text{=12, N}_{\text{UE}}\text{=4, Sim}$}

\addplot [color=blue, line width=2.0pt, draw=none, mark=triangle, mark options={solid, rotate=270, blue}]
  table[row sep=crcr]{%
0	0.126198443035535\\
0.005	0.104348715596782\\
0.01	0.290661105417303\\
0.015	0.472352210780704\\
0.02	0.605845579065449\\
0.025	0.698836247424504\\
0.03	0.764310093798944\\
0.035	0.811583264798097\\
0.04	0.846604110971308\\
0.045	0.873149448908951\\
0.05	0.893672452784401\\
0.055	0.90981243639234\\
0.06	0.922695577686989\\
0.065	0.933114375245052\\
0.07	0.94163939150908\\
0.075	0.948688599847234\\
0.08	0.95457342391669\\
0.085	0.959529302225574\\
0.09	0.963736465026759\\
0.095	0.967334584047568\\
0.1	0.970432967618783\\
};
\addlegendentry{$\text{N}_{\text{BS}}\text{=12, N}_{\text{UE}}\text{=4, Analysis}$}

\addplot [color=mycolor1, dashdotted, line width=2.0pt]
  table[row sep=crcr]{%
0	0.4321\\
0.005	0.3878\\
0.01	0.4727\\
0.015	0.5852\\
0.02	0.663\\
0.025	0.7237\\
0.03	0.7771\\
0.035	0.8163\\
0.04	0.8437\\
0.045	0.8721\\
0.05	0.8837\\
0.055	0.9026\\
0.06	0.9164\\
0.065	0.93\\
0.07	0.9324\\
0.075	0.9396\\
0.08	0.9477\\
0.085	0.9535\\
0.09	0.9563\\
0.095	0.96\\
0.1	0.9654\\
};
\addlegendentry{$\text{N}_{\text{BS}}\text{=3, N}_{\text{UE}}\text{=4, Sim}$}

\addplot [color=mycolor2, line width=2.0pt, draw=none, mark=square, mark options={solid, mycolor2}]
  table[row sep=crcr]{%
0	0.537931510401758\\
0.005	0.339812109238662\\
0.01	0.426305873148247\\
0.015	0.541738374331302\\
0.02	0.643243594436266\\
0.025	0.722610019360015\\
0.03	0.782263190043683\\
0.035	0.826715662101673\\
0.04	0.860006897357534\\
0.045	0.885216292267737\\
0.05	0.904565866981717\\
0.055	0.919631095397308\\
0.06	0.931526618227263\\
0.065	0.941046591573474\\
0.07	0.948762108100073\\
0.075	0.955088874149438\\
0.08	0.960333158794422\\
0.085	0.964702377411939\\
0.09	0.968432530491133\\
0.095	0.971592210055501\\
0.1	0.974304568537176\\
};
\addlegendentry{$\text{N}_{\text{BS}}\text{=3, N}_{\text{UE}}\text{=4, Analysis}$}

\addplot [color=red, dashed, line width=2.0pt]
  table[row sep=crcr]{%
0	0.7324\\
0.005	0.638\\
0.01	0.607\\
0.015	0.6182\\
0.02	0.6437\\
0.025	0.6814\\
0.03	0.7165\\
0.035	0.7527\\
0.04	0.7793\\
0.045	0.8053\\
0.05	0.8332\\
0.055	0.8487\\
0.06	0.8677\\
0.065	0.8832\\
0.07	0.8965\\
0.075	0.9089\\
0.08	0.9163\\
0.085	0.9248\\
0.09	0.9308\\
0.095	0.937\\
0.1	0.9419\\
};
\addlegendentry{$\text{N}_{\text{BS}}\text{=1, N}_{\text{UE}}\text{=1, Sim}$}

\addplot [color=red, line width=2.0pt, draw=none, mark=o, mark options={solid, red}]
  table[row sep=crcr]{%
0	0.789761724808978\\
0.005	0.64165155050797\\
0.01	0.604251891532881\\
0.015	0.611615140308837\\
0.02	0.641565721062251\\
0.025	0.679544671544794\\
0.03	0.717920645170956\\
0.035	0.753325862181877\\
0.04	0.784609511160662\\
0.045	0.811664100410032\\
0.05	0.834822408972667\\
0.055	0.854565739399209\\
0.06	0.871390606741341\\
0.065	0.885753174542478\\
0.07	0.898050652638792\\
0.075	0.908619478427654\\
0.08	0.917740468942924\\
0.085	0.925646285594094\\
0.09	0.932529095498568\\
0.095	0.938547551981491\\
0.1	0.943832806731941\\
};
\addlegendentry{$\text{N}_{\text{BS}}\text{=1, N}_{\text{UE}}\text{=1, Analysis}$}

\end{axis}

\begin{axis}[%
width=0.7\columnwidth,
height=0.4\columnwidth,
at={(0\columnwidth,0\columnwidth)},
scale only axis,
xmin=0,
xmax=1,
ymin=0,
ymax=1,
axis line style={draw=none},
ticks=none,
axis x line*=bottom,
axis y line*=left,
legend style={legend cell align=left, align=left, draw=white!15!black}
]
\end{axis}
\end{tikzpicture}%
}

\caption{Effect of blockage on detection failure probability, $\lambda = 10^{-4}$.}
\label{fig:blockage}
\end{figure}
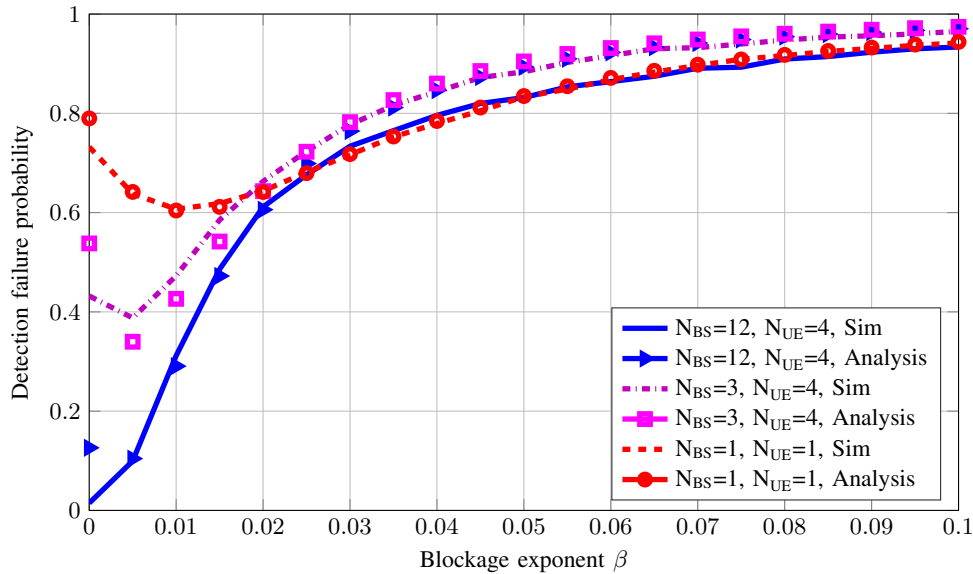

Fig.~\ref{fig:blockage} illustrates the effect of blockage on failure probability with different beamwidths. For narrow (mainlobe) beams, the failure probability increases with the obstacle density because most available BSs are blocked. When the (mainlobe) beamwidth is large, there are more BSs pointing to the typical UE with mainlobe during one slot, which increases the received interference. Altogether, the growing obstacle density first reduces the interference, thereby improving the performance at the beginning. After a critical point of the obstacle density, the failure probability starts to increase as the former scenario. Also, we observe that the failure probabilities of all beamwidths converge to a similar value in a densely blocked scenario, like an office room. Thus, we may even utilize wider beams in mmWave since the effective transmission distance is very short.

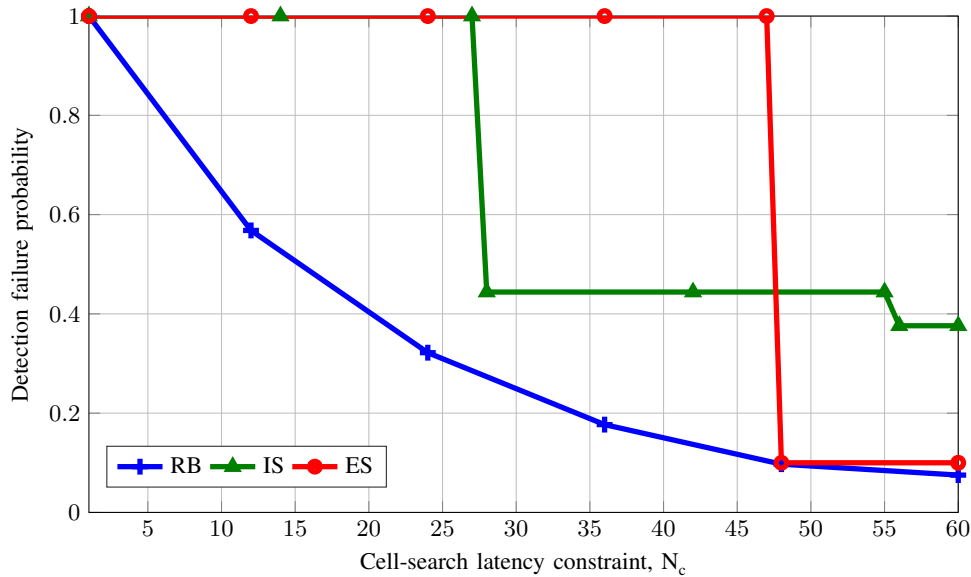
\begin{figure}
\centering
\footnotesize{
%
%
\definecolor{mycolor1}{rgb}{0.00000,0.49804,0.00000}%
\begin{tikzpicture}

\begin{axis}[%
width=0.7\columnwidth,
height=0.4\columnwidth,
at={(0\columnwidth,0\columnwidth)},
scale only axis,
xmin=1,
xmax=60,
xlabel={$\text{Cell-search latency constraint, N}_\text{c}$},
ymin=0,
ymax=1,
ylabel={Detection failure probability},
axis background/.style={fill=white},
xmajorgrids,
ymajorgrids,
legend style={at={(0.02,0.05)}, anchor=south west, legend cell align=left, align=left,draw=white!15!black,legend columns=-1}
]
\addplot [color=blue, line width=2.0pt, mark size=3.0pt, mark=+, mark options={solid, blue}]
  table[row sep=crcr]{%
1	1\\
12	0.5681\\
24	0.3218\\
36	0.1769\\
48	0.0972\\
60	0.0752\\
};
\addlegendentry{RB}

\addplot [color=mycolor1, line width=2.0pt, mark=triangle, mark options={solid, mycolor1}]
  table[row sep=crcr]{%
1	1\\
14	1\\
27	1\\
28	0.444\\
42	0.444\\
55	0.444\\
56	0.3761\\
60	0.3761\\
};
\addlegendentry{IS}

\addplot [color=red, line width=2.0pt, mark=o, mark options={solid, red}]
  table[row sep=crcr]{%
1	1\\
12	1\\
24	1\\
36	1\\
47	1\\
48	0.1\\
60	0.1\\
};
\addlegendentry{ES}

\end{axis}

\begin{axis}[%
width=0.7\columnwidth,
height=0.4\columnwidth,
at={(0\columnwidth,0\columnwidth)},
scale only axis,
xmin=0,
xmax=1,
ymin=0,
ymax=1,
axis line style={draw=none},
ticks=none,
axis x line*=bottom,
axis y line*=left,
legend style={legend cell align=left, align=left, draw=white!15!black}
]
\end{axis}
\end{tikzpicture}%
}
\caption{Tradeoff between the detection failure probability and average latency, $\lambda = 10^{-4}$. }
\label{fig:dfpcomp}
\end{figure}

Fig.~\ref{fig:dfpcomp} illustrates the tradeoff between detection failure probability and the cell-search budget. The areas above the lines illustrate the feasible performance regions of cell-search, namely there are some settings for $\Nb$ and $\Nu$ that allow realizing any $(\mathbb{E}[D_I], P_f)$ point above the line. For the iterative search and exhaustive search, the feasible regions are sequences of step functions due to their quantized latency. Moreover, denser mmWave networks have a larger feasible region, which gives more flexibility to optimize the tradeoff between detection probability and latency.


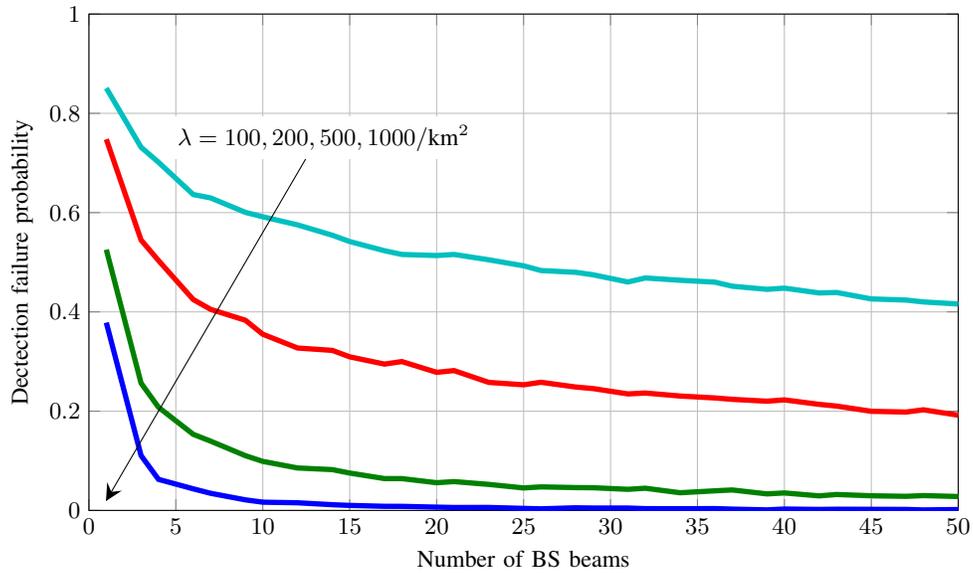
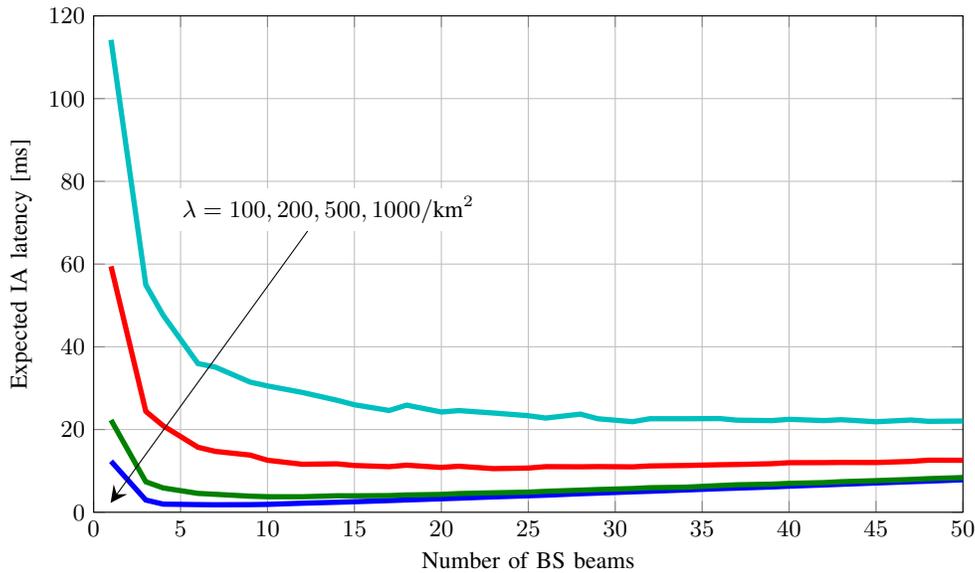
\begin{figure}[t]
\centering
\begin{subfigure}[a]{\textwidth}
 \centering
\footnotesize{
%
%
\definecolor{mycolor1}{rgb}{0.00000,0.75000,0.75000}%
\begin{tikzpicture}

\begin{axis}[%
width=0.7\columnwidth,
height=0.4\columnwidth,
at={(0\columnwidth,0\columnwidth)},
scale only axis,
xmin=0,
xmax=50,
xlabel={Number of BS beams},
ymin=0,
ymax=1,
ylabel={Dectection failure probability},
axis background/.style={fill=white},
xmajorgrids,
ymajorgrids,
legend style={legend cell align=left, align=left, draw=white!15!black}
]
\addplot [color=blue, line width=2.0pt]
  table[row sep=crcr]{%
1	0.3784\\
3	0.1104\\
4	0.0626\\
6	0.0435\\
7	0.0345\\
9	0.0213\\
10	0.0166\\
12	0.0154\\
14	0.0115\\
15	0.0101\\
17	0.0083\\
18	0.0082\\
20	0.0066\\
21	0.0059\\
23	0.006\\
25	0.0041\\
26	0.0032\\
28	0.0053\\
29	0.0048\\
31	0.0048\\
32	0.0036\\
34	0.0034\\
36	0.0037\\
37	0.0026\\
39	0.0011\\
40	0.0028\\
42	0.002\\
43	0.0024\\
45	0.0022\\
47	0.002\\
48	0.001\\
50	0.0018\\
};

\addplot [color=black!50!green, line width=2.0pt]
  table[row sep=crcr]{%
1	0.5254\\
3	0.2564\\
4	0.2079\\
6	0.1533\\
7	0.1396\\
9	0.1103\\
10	0.0988\\
12	0.0855\\
14	0.0822\\
15	0.0753\\
17	0.0641\\
18	0.0641\\
20	0.0558\\
21	0.0581\\
23	0.0524\\
25	0.0453\\
26	0.0474\\
28	0.046\\
29	0.0457\\
31	0.0425\\
32	0.0447\\
34	0.0356\\
36	0.0395\\
37	0.0412\\
39	0.0333\\
40	0.0352\\
42	0.0293\\
43	0.0321\\
45	0.0295\\
47	0.0284\\
48	0.0299\\
50	0.028\\
};

\addplot [color=red, line width=2.0pt]
  table[row sep=crcr]{%
1	0.7478\\
3	0.5447\\
4	0.5035\\
6	0.4249\\
7	0.4051\\
9	0.3831\\
10	0.355\\
12	0.3272\\
14	0.3224\\
15	0.3094\\
17	0.2948\\
18	0.2999\\
20	0.2782\\
21	0.2817\\
23	0.2579\\
25	0.2531\\
26	0.2582\\
28	0.2486\\
29	0.2455\\
31	0.2347\\
32	0.2365\\
34	0.2305\\
36	0.2266\\
37	0.2238\\
39	0.22\\
40	0.2227\\
42	0.2137\\
43	0.2102\\
45	0.1997\\
47	0.1982\\
48	0.2027\\
50	0.1919\\
};

\addplot [color=mycolor1, line width=2.0pt]
  table[row sep=crcr]{%
1	0.8508\\
3	0.7316\\
4	0.7015\\
6	0.6364\\
7	0.6296\\
9	0.6005\\
10	0.5916\\
12	0.5753\\
14	0.5544\\
15	0.5417\\
17	0.5232\\
18	0.5157\\
20	0.5135\\
21	0.5157\\
23	0.5049\\
25	0.4928\\
26	0.4833\\
28	0.4797\\
29	0.4747\\
31	0.4602\\
32	0.4683\\
34	0.4637\\
36	0.4599\\
37	0.4517\\
39	0.4454\\
40	0.4478\\
42	0.4381\\
43	0.4391\\
45	0.4261\\
47	0.4238\\
48	0.42\\
50	0.4159\\
};

\end{axis}

\begin{axis}[%
width=0.7\columnwidth,
height=0.4\columnwidth,
at={(0\columnwidth,0\columnwidth)},
scale only axis,
xmin=0,
xmax=1,
ymin=0,
ymax=1,
axis line style={draw=none},
ticks=none,
axis x line*=bottom,
axis y line*=left,
legend style={legend cell align=left, align=left, draw=white!15!black}
]
\draw [decoration={markings,mark=at position 1 with
    {\arrow[scale=2,>=stealth]{>}}},postaction={decorate},black] (0.26,0.74) -- (0.02,0.02);
\node[draw=none,fill=white] at (0.27,0.75) {$\lambda=100,200,500,1000 / \text{km}^2$};
\end{axis}
\end{tikzpicture}%
}
\caption{Detection failure probability vs number of BS beams.}
\label{fig:beamwidthpf}
\end{subfigure}
\begin{subfigure}[a]{\textwidth}
\centering
\footnotesize{
%
%
\definecolor{mycolor1}{rgb}{0.00000,0.75000,0.75000}%
\begin{tikzpicture}

\begin{axis}[%
width=0.7\columnwidth,
height=0.4\columnwidth,
at={(0\columnwidth,0\columnwidth)},
scale only axis,
xmin=0,
xmax=50,
xlabel={Number of BS beams},
ymin=0,
ymax=120,
ylabel={Expected IA latency [ms]},
axis background/.style={fill=white},
xmajorgrids,
ymajorgrids,
legend style={legend cell align=left, align=left, draw=white!15!black}
]
\addplot [color=blue, line width=2.0pt]
  table[row sep=crcr]{%
1	12.3312821750322\\
3	2.95076438848921\\
4	1.96060913164071\\
6	1.84706612650287\\
7	1.80840561885034\\
9	1.84152127822622\\
10	1.90010423022168\\
12	2.18781738777168\\
14	2.42017577137076\\
15	2.54781101626427\\
17	2.82363933145104\\
18	2.97785591853196\\
20	3.2578769881216\\
21	3.39995033195856\\
23	3.71447434607646\\
25	3.98858758409479\\
26	4.12670545746388\\
28	4.48156479340505\\
29	4.62771302250804\\
31	4.94021302250804\\
32	5.07226013649137\\
34	5.38073198876179\\
36	5.69927481682224\\
37	5.83338555243633\\
39	6.11577422664931\\
40	6.30615724027276\\
42	6.60258016032064\\
43	6.76686547714515\\
45	7.07534701342955\\
47	7.38383016032064\\
48	7.52002002002002\\
50	7.84856491685033\\
};

\addplot [color=black!50!green, line width=2.0pt]
  table[row sep=crcr]{%
1	22.2970001053519\\
3	7.36493074233459\\
4	5.87433720489837\\
6	4.55861727884729\\
7	4.33875232450023\\
9	3.88573746768574\\
10	3.75513204616067\\
12	3.74487424822307\\
14	3.97873992155154\\
15	3.97238631448037\\
17	4.02605446628913\\
18	4.18230446628913\\
20	4.30695297606439\\
21	4.51492661110521\\
23	4.69970187842972\\
25	4.85523921127056\\
26	5.05767111064455\\
28	5.3393605870021\\
29	5.48902009326208\\
31	5.73147845953003\\
32	5.93583167591333\\
34	6.05078287017835\\
36	6.44748828735034\\
37	6.64065759282436\\
39	6.78269176062894\\
40	6.97968490878939\\
42	7.16618806016277\\
43	7.38204166236181\\
45	7.63918405461103\\
47	7.92835271716756\\
48	8.11643129574271\\
50	8.3886316872428\\
};

\addplot [color=red, line width=2.0pt]
  table[row sep=crcr]{%
1	59.4583911578113\\
3	24.3958310454645\\
4	20.906973816717\\
6	15.7140605981568\\
7	14.7128456463271\\
9	13.8264153428432\\
10	12.5702519379845\\
12	11.6015160523187\\
14	11.7034386068477\\
15	11.3040743556328\\
17	11.0169987237663\\
18	11.3798475217826\\
20	10.833506511499\\
21	11.1247694208548\\
23	10.5442957485514\\
25	10.6835963649752\\
26	11.0239451334592\\
28	10.9919816342827\\
29	11.0388709410205\\
31	10.9772924016725\\
32	11.1951538965291\\
34	11.3034031838856\\
36	11.4848396689941\\
37	11.5478050115949\\
39	11.7347756410256\\
40	11.9800913418243\\
42	11.9980843825512\\
43	12.0416165484933\\
45	12.0218785143071\\
47	12.2876262783737\\
48	12.5846607299636\\
50	12.5619122014602\\
};

\addplot [color=mycolor1, line width=2.0pt]
  table[row sep=crcr]{%
1	114.204507372654\\
3	54.9843982861401\\
4	47.626675041876\\
6	35.943000550055\\
7	35.0894303455724\\
9	31.4688282227785\\
10	30.5340964740451\\
12	28.9670649870497\\
14	27.0708034111311\\
15	25.9832874209033\\
17	24.6025587248322\\
18	25.8880976739177\\
20	24.2349691675231\\
21	24.5779669110056\\
23	23.9896296202787\\
25	23.3384266561514\\
26	22.7696801819237\\
28	23.7218424158961\\
29	22.6047318199124\\
31	21.8945095405706\\
32	22.6151965394019\\
34	22.6050601342532\\
36	22.655179596371\\
37	22.2576315429509\\
39	22.1557766858997\\
40	22.4687613183629\\
42	22.1560219789998\\
43	22.3757298538064\\
45	21.8805268775048\\
47	22.296563701503\\
48	21.9827586206897\\
50	22.0532122068139\\
};

\end{axis}

\begin{axis}[%
width=0.7\columnwidth,
height=0.4\columnwidth,
at={(0\columnwidth,0\columnwidth)},
scale only axis,
xmin=0,
xmax=1,
ymin=0,
ymax=1,
axis line style={draw=none},
ticks=none,
axis x line*=bottom,
axis y line*=left,
legend style={legend cell align=left, align=left, draw=white!15!black}
]
\draw [decoration={markings,mark=at position 1 with
    {\arrow[scale=2,>=stealth]{>}}},postaction={decorate},black] (0.26,0.6) -- (0.02,0.02);
\node[draw=none,fill=white] at (0.27,0.61) {$\lambda=100,200,500,1000 / \text{km}^2$};
\end{axis}
\end{tikzpicture}%
}
\caption{Expected IA latency vs number of BS beams.}
\label{fig:beamwidthdelay}
\end{subfigure}
\caption{Effect of BS sectors on the performance metrics.}
\label{fig:beamwidth}
\end{figure}

\begin{figure}
\centering
\footnotesize{
%
%
\begin{tikzpicture}

\begin{axis}[%
width=0.7\columnwidth,
height=0.4\columnwidth,
at={(0\columnwidth,0\columnwidth)},
scale only axis,
xmode=log,
xmin=999,
xmax=1000000000,
xminorticks=true,
xlabel={Packet size L [bits]},
ymode=log,
ymin=1,
ymax=10000,
yminorticks=true,
xmajorgrids,
ymajorgrids,
ylabel={Total data transmission latency [ms]},
axis background/.style={fill=white},
legend style={at={(0.02,0.98)}, anchor=north west, legend cell align=left, align=left,draw=white!15!black,legend columns=-1}
]
\addplot [color=blue, line width=2.0pt, mark size=3.0pt, mark=+, mark options={solid, blue}]
  table[row sep=crcr]{%
1000	27.9578349282297\\
10000	27.9874401913876\\
100000	28.2834928229665\\
1000000	31.244019138756\\
10000000	63.3492822966507\\
100000000	401.901913875598\\
1000000000	3784.92822966507\\
};
\addlegendentry{RB}

\addplot [color=black!50!green, line width=2.0pt, mark=triangle, mark options={solid, black!50!green}]
  table[row sep=crcr]{%
1000	19.7233063549161\\
10000	19.7420563549161\\
100000	19.9295563549161\\
1000000	21.8045563549161\\
10000000	44.3045563549161\\
100000000	273.054556354916\\
1000000000	2583.05455635492\\
};
\addlegendentry{IS}

\addplot [color=red, line width=2.0pt, mark=o, mark options={solid, red}]
  table[row sep=crcr]{%
1000	8.47456414259693\\
10000	8.49564142596929\\
100000	8.70641425969295\\
1000000	10.8141425969295\\
10000000	38.1414259692948\\
100000000	348.914259692948\\
1000000000	3412.89259692948\\
};
\addlegendentry{ES}

\addplot [color=blue, line width=2.0pt, mark size=3.0pt, mark=+, mark options={solid, blue}, forget plot]
  table[row sep=crcr]{%
1000	2.82773489760214\\
10000	2.85051970772872\\
100000	3.07836780899455\\
1000000	5.35684882165277\\
10000000	30.6416589482351\\
100000000	290.989760214058\\
1000000000	2894.47077287229\\
};
\addplot [color=black!50!green, line width=2.0pt, mark=triangle, mark options={solid, black!50!green}, forget plot]
  table[row sep=crcr]{%
1000	3.94125455754682\\
10000	3.95448985166447\\
100000	4.08684279284094\\
1000000	5.41037220460565\\
10000000	18.6456663222527\\
100000000	184.748607498723\\
1000000000	1812.02801926343\\
};
\addplot [color=red, line width=2.0pt, mark=o, mark options={solid, red}, forget plot]
  table[row sep=crcr]{%
1000	6.25135135135135\\
10000	6.26351351351351\\
100000	6.38513513513514\\
1000000	7.60135135135135\\
10000000	19.7635135135135\\
100000000	197.635135135135\\
1000000000	1970.10135135135\\
};
\end{axis}

\begin{axis}[%
width=0.7\columnwidth,
height=0.4\columnwidth,
at={(0\columnwidth,0\columnwidth)},
scale only axis,
xmin=0,
xmax=1,
ymin=0,
ymax=1,
axis line style={draw=none},
ticks=none,
axis x line*=bottom,
axis y line*=left,
legend style={legend cell align=left, align=left, draw=white!15!black}
]
\draw [black, line width=1.0pt] (axis cs:0.474607,0.18) ellipse [x radius=0.013, y radius=0.06];
\draw [black, line width=1.0pt] (axis cs:0.568586,0.36) ellipse [x radius=0.014, y radius=0.07];
\draw [decoration={markings,mark=at position 1 with
    {\arrow[scale=2,>=stealth]{>}}},postaction={decorate},black] (0.7,0.1) -- (0.5,0.14);
\node[draw=none,fill=white] at (0.73,0.11) {$\lambda = 10^{-3}$};
\draw [decoration={markings,mark=at position 1 with
    {\arrow[scale=2,>=stealth]{>}}},postaction={decorate},black] (0.43,0.53) -- (0.54,0.42);
\node[draw=none,fill=white] at (0.4,0.55) {$\lambda = 10^{-4}$};
\end{axis}
\end{tikzpicture}%
}
\caption{Total data transmission latency vs packet size.}
\label{fig:totaldelay}
\end{figure}
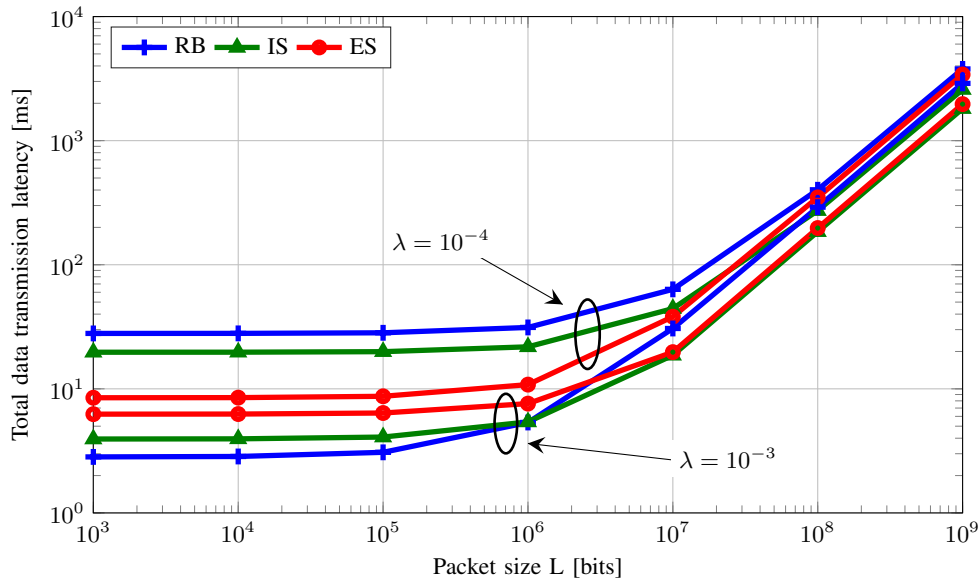

Fig.~\ref{fig:beamwidth} shows the objective and constraint functions of optimization problem~\eqref{eq: optimization-problem}. To perform a fair comparison, we make a small adaption to the frame structure and set the length of SS burst to the duration of one scan cycle for different BS beamwidth values. As shown in Fig.~\ref{fig:beamwidthpf}, the detection failure probability is a decreasing function of $\Nb$ and therefore narrower beams are always beneficial. The floor of $P_f$ is due to both the blockage and deep fading, which can be improved either by increasing the BS density ($\lambda$) or search budget ($N_c$). Meanwhile, the IA delay drops as well until some point. After the critical point where the decrease in $P_f$ is negligible, the IA delay increases with $\Nb$ due to longer overhead.
In this situation, narrower beams cannot reduce the number of extra required frames anymore but almost linearly increases the search latency. Therefore, the expected IA latency shows a convex shape and the minimum value is shifting depending on BS density. The optimal beamwidth (or equivalently the optimal $\Nb$) depends on the maximum allowable detection failure probability $P_f^{\text{max}}$ and the deployment situation of the network.

At last, we present the result of total data transmission delay in Fig.~\ref{fig:totaldelay}. When the BS density is not large enough ($\lambda = 10^{-4}$), other schemes always outperform random beamforming due to its high failure probability and low achievable rate. However in dense regime ($\lambda = 10^{-3}$), as the available BSs increase, we achieve a lower total delay for relative shorter packet sizes with IA under random beamforming. Thus the most favorable scenario for IA under random beamforming is transmitting short packets in dense networks.

\section{Conclusion and Future Works}\label{sec: conclusions}
In this paper, we investigated the performance of random beamforming in initial cell-search of mmWave networks. We developed an analytical framework leveraging stochastic geometry to evaluate the detection failure and latency performance. We compared our method with two sophisticated schemes in both control and data plane. The numerical results showed that random beamforming, while being very efficient from signaling and computational perspectives, can provide sufficient detection and latency performance in control plane, especially in dense BS deployment scenarios. The optimal beamwidth to achieve a minimum search latency depends on the detection requirement and BS density. From data plane perspective, random beamforming outperforms other schemes for short packet transmission in high density regimes. Consequently, it may be unnecessary to develop complex algorithms for cell search process in future dense mmWave networks. Meanwhile, random beamforming scheme can be selected as a new benchmark in future initial access studies considering its simplicity and good performance.

This paper has shown that random beamforming is effective for dense mmWave networks in an average sense. Nevertheless, each specific cell may benefit from different beamwidths and/or beam patterns. The recent success of deep learning underpins new and powerful tools that may help manage such situations. For future work, it would be interesting to develop a learning based initial access scheme serving every BS in the mmWave network. A learning algorithm may generate individual cell-search antenna patterns as well as random access schemes for each BS that further reduces the IA latency and increases the transmission data rate.
\appendices
\section*{Appendix\\Proofs}

\subsection{Proposition~\ref{prop: failure-prob-mainlobe}}
\label{sec:P1}
Given the antenna directivity, we can consider the BSs pointing to the typical UE as a thinning process $\Phi(\lambda^{\prime})$ from the original process $\Phi(\lambda)$ within one mini-slot, where the effective density $\lambda^{\prime} = \frac{\Thetab}{2\pi}\lambda$. Furthermore, since we consider BS beamwidth smaller than $\frac{\pi}{2}$, the BS and UE beams will be aligned for only 1 mini-slot during 1 scanning cycle. Thus the process $\Phi(\lambda^{\prime})$ can be divided into $\Nb$ independent PPPs $\Phi_m(\lambda^{\prime})$, $m=1,\dots, \Nb$, where $\Phi_m(\lambda^{\prime})$ consists of the BSs pointing to the typical UE with mainlobe in the $m$-th mini-slot of the scan cycle. Therefore, we can treat each mini-slot independently and write the detection failure probability after $N_c$ mini-slots as $(1-P_s)^{N_c}$, where $P_s$ denotes the detection success probability of 1 mini-slot. 

Let $\text{BS}_{i}$ denote the BS located at $x_i$, $r = \lVert x_i \rVert$ and $I_{x_i} \triangleq \sum\limits_{x_j \in \Phi \backslash x_i} h_{j} \ell(\lVert x_j \rVert)  S_{j}$. Then, the successful detection probability in one mini-slot under strongest BS association can be derived as follows:
\begin{align}
\label{equ:ps1}
P_s &= \Pr\(\max\limits_{x_i \in \Phi} \text{SINR}_{x_i} \geq T\) = \Pr\(\bigcup_{x_i \in \Phi} \text{SINR}_{i} \geq T \) \nonumber \\
&\stackrel{(a)}{=} \mathbb{E} \[\sum\limits_{x_i \in \Phi} \mathbbm{1}(\text{SINR}_{i}\geq T)\] \nonumber\\
&\stackrel{(b)}{=} \frac{\Thetab}{2 \pi} \lambda \int_{\mathbb{R}^2} \Pr(\text{SINR}_{i} \geq T \mid r) \mathrm{d}r \nonumber\\
&= \frac{\Thetau\Thetab}{2 \pi} \lambda \int_{0}^{\infty} \hspace{-2.5mm} \Pr(\text{SINR}_{i} \geq T \hspace{-0.8mm} \mid \hspace{-0.8mm} S_i=1, r) \hspace{-0.8mm} \Pr(S_i=1 \mid r) r\mathrm{d}r \nonumber\\
&\stackrel{(c)}{=}  \frac{2 \pi}{\Nb \Nu} \lambda \int_{0}^{\infty} \mathcal{L}_{I_{x_i}} (T r^{\alpha}) e^{-T r^{\alpha} \sigma^2} e^{-\beta r} r\mathrm{d}r \:,
\end{align}
where (a) follows from Lemma 1 in \cite{Dhillon2012} \begin{footnote}{Note that Lemma 1 in \cite{Dhillon2012} is based on $T>1$ (0dB). It also provides a tight upper bound until $T=0.4$ (-4dB).}\end{footnote}, (b) follows from Campbell Mecke Theorem \cite{Chiu2013} and (c) follows the Rayleigh fading assumption. The use of $\Thetau$ and $\Thetab\lambda/2\pi$ are due to BS and UE beamwidth. Here $\mathcal{L}_{I_{x_i}}(T r^{\alpha})$ is the Laplace transform of the interference $I_{x_i}$. Letting $R_j$ denote the distance from the $j$th interfering BS to the typical UE, $\mathcal{L}_{I_{x_i}}(T r^{\alpha})$ can be expressed as:

\begin{align}\label{eq:laplace1}
\mathcal{L}_{I_{x_i}} & (T r^{\alpha}) = \mathbb{E}_{\Phi, h_i}\[\exp \(-T r^{\alpha} \sum\limits_{x_j \in \Phi \backslash x_i} \ell(R_j) h_j S_j \) \]  \nonumber \\
&\stackrel{(a)}{=} \mathbb{E}\[\prod\limits_{x_j \in \Phi \backslash x_i} \hspace{-3mm} \mathbb{E}_{h_j}\[\exp(-T r^{\alpha} R_j^{-\alpha} h_j)\]e^{-\beta R_j} + 1 - e^{\beta R_j} \] \nonumber \\[-0.5mm]
&\stackrel{(b)}{=} \mathbb{E}\[\prod\limits_{x_j \in \Phi \backslash x_i} 1-\frac{T r^{\alpha}e^{-\beta R_j}}{R_j^{\alpha}+T r^{\alpha}}\]  \nonumber \\
&\stackrel{(c)}{=} \exp\(-\Thetau\frac{\Thetab}{2\pi} \lambda \int_0^{\infty} \frac{T r^{\alpha}e^{-\beta v}}{v^{\alpha}+T r^{\alpha}} v\mathrm{d}v\) \:,
\end{align}
where (a) follows that $S_j$ is a Bernoulli random variable with parameter $e^{-\beta R_i}$, (b) follows that $h_j$ is an exponential random variable and (c) is derived from the probability generating function of the PPP. Substituting (\ref{eq:laplace1}) into \eqref{equ:ps1} we obtain the successful detection probability in one mini-slot. 



\subsection{Lemma~\ref{lem: laplace-nlos}}
When $\alpha_N<\infty$, the interference is composed of LOS part and NLOS part. Similar as in \ref{eq:laplace1}, $R_j$ denotes the distance from the $j$th interfering BS to the typical UE and  $\mathcal{L}_{I}^N(s)$ can be expressed as:

\begin{align}\label{eq:laplace2}
\mathcal{L}_{I}^N (s) &= \mathbb{E}_{I_{x_i}} [e^{-s I_{x_i}}]= \mathbb{E}_{\Phi, h_i}\[\exp \(-s \sum\limits_{x_j \in \Phi \backslash x_i} \ell(R_{j}) h_j S_j \) \]  \nonumber \\
&= \mathbb{E}\[\prod\limits_{x_j \in \Phi \backslash x_i} \mathbb{E}_{\alpha_j, h_j}\[\exp(-s (k R_j)^{-\alpha_j}  h_j S_j\]\]  \nonumber \\
&\stackrel{(a)}{=} \mathbb{E}\[\prod\limits_{x_j \in \Phi \backslash x_i} \hspace{-3mm} \mathbb{E}_{h_j}\[\exp(-s (k R_j)^{-\alpha_L} h_j) e^{-\beta R_j} + \exp(-s (k R_j)^{-\alpha_N} h_j) (1 - e^{\beta R_j})\] \] \nonumber \\[-0.5mm]
&\stackrel{(b)}{=} \mathbb{E}\[\prod\limits_{x_j \in \Phi \backslash x_i} \frac{R_j^{\alpha_L} e^{-\beta R_j}}{R_j^{\alpha_L}+s k_1} + \frac{R_j^{\alpha_N} (1-e^{-\beta R_j})}{R_j^{\alpha_N}+s k_2}\]  \nonumber \\
&\stackrel{(c)}{=} \exp\(-\frac{2\pi}{\Nb \Nu} \lambda \int_{0}^{\infty}(1-\frac{e^{-\beta v} v^{\alpha_L}}{v^{\alpha_L}+s k_1} - \frac{(1-e^{-\beta v}) v^{\alpha_N}}{v^{\alpha_N}+s k_2})v\mathrm{d}v\) \:,
\end{align}
where (a) follows LOS/NLOS conditions with probability $e^{-\beta R_j}$, (b) follows that $h_j$ is an exponential random variable and (c) is derived from the probability generating function of the PPP. 

\subsection{Proposition~\ref{prop: failure-prob-NLoS}}
\label{sec:P2}
Similar to Proposition \ref{prop: failure-prob-mainlobe}, the successful detection probability in one mini-slot under strongest BS association can be derived as follows:

\begin{align}\label{Ps}
P_s^N &={} \frac{\Thetab}{2 \pi} \lambda \int_{\mathbb{R}^2} \Pr(\text{SINR}_{i} \geq T \mid r) \mathrm{d}r \nonumber \\
\begin{split}
&={} \frac{\Thetau\Thetab}{2 \pi} \lambda \int_{0}^{\infty} ( \Pr(\text{SINR}_{i} \geq T \mid S_i=1, r) \Pr(S_i=1 \mid r) \nonumber \\
& \hspace{4cm} + \Pr(\text{SINR}_{i} \geq T \mid S_i=0, r) \Pr(S_i=0 \mid r) )r\mathrm{d}r \\
\end{split}\\
\begin{split}
&={} \frac{2 \pi}{\Nb \Nu} \lambda \int_{0}^{\infty} \left( e^{-T C(r)^{-\alpha_L} \sigma^2} \calL_{I_r}(T C(r)^{-\alpha_L}) e^{-\beta r}  \right.\\
& \hspace{4cm} \left. + e^{-T C(r)^{-\alpha_N} \sigma^2} \calL_{I_r}(T C(r)^{-\alpha_N}) (1-e^{-\beta r}) \right) r\mathrm{d}r \:,
\end{split}
\end{align}
where the two steps follow from Campbell Mecke Theorem the Rayleigh fading assumption as in Proposition~\ref{prop: failure-prob-mainlobe}. 

\subsection{Lemma~\ref{lem: mainlobe-success}}
We start from deriving the Laplace transform of the interference which comprises the mainlobe tier and sidelobe tier: 

\begin{align}\label{eq:laplace-mainlobe1}
\mathcal{L}_{I_{x_i}}^m (s) &= \prod\limits_{j=1}^{2} \mathbb{E}_{\Phi}\[\prod\limits_{x_j \in \Phi_j \backslash x_i} \mathbb{E}_{h_j}\[\exp(-s G_{\text{BS}_j} R_j^{-\alpha}  h_j \]\]  \nonumber \\
&\stackrel{(a)}{=} \prod\limits_{j=1}^{2} \mathbb{E}_{\Phi}\[\prod\limits_{x_j \in \Phi \backslash x_i} \hspace{-3mm} \mathbb{E}_{h_j}\[\exp(-s G_{\text{BS}_j} R_j^{-\alpha} h_j) e^{-\beta R_j} + 1 - e^{\beta R_j}\] \] \nonumber \\[-0.5mm]
&\stackrel{(b)}{=} \prod\limits_{j=1}^{2} \mathbb{E}\[\prod\limits_{x_j \in \Phi \backslash x_i} 1- \frac{s P_j R_j^{-\alpha} e^{-\beta R_j}}{1+s G_{\text{BS}_j} R_j^{-\alpha}} \]  \nonumber \\
&= \prod\limits_{j=1}^{2} \exp\(-\Thetau \lambda_j \int_{0}^{\infty}\frac{s G_{\text{BS}_j}  e^{-\beta R_j}}{v^{-\alpha}+s G_{\text{BS}_j} }v\mathrm{d}v\) \:,
\end{align}
where (a) follows the blockage model with parameter $e^{-\beta R_j}$, (b) follows that $h_j$ is an exponential random variable and $\lambda_j = \{ \frac{\Thetab}{2 \pi} \lambda,  \frac{2 \pi - \Thetab}{2 \pi} \lambda\}$. Putting $s=\frac{Tr^\alpha}{G_{\text{BS}_i}}$, we have
\begin{align}\label{eq:laplace-mainlobe2}
\mathcal{L}_{I_{x_i}}^m (\frac{Tr^\alpha}{G_{\text{BS}_i}}) = \prod\limits_{j=1}^{2} \exp\(-\Thetau \lambda_j \int_{0}^{\infty}\frac{T r^{\alpha} G_{\text{BS}_j}  e^{-\beta R_j}}{v^{-\alpha} G_{\text{BS}_i} + T r^{\alpha} G_{\text{BS}_j} }v\mathrm{d}v\)
\end{align}

Then, similar as in Proposition~\ref{prop: failure-prob-mainlobe}, the successful detection probability in one mini-slot under strongest BS association can be derived as follows:

\begin{align}
\label{equ:Ps}
P_{\text{sm}} &= \Pr\(\bigcup_{x_i \in \Phi_m} \text{SINR}_{x_i} \geq T \) \nonumber \\
&= \lambda_i \int_{\mathbb{R}^2} \Pr(\text{SINR}_{x_i} \geq T \mid r) \mathrm{d}r \nonumber\\
&= \Thetau \lambda_i \int_{0}^{\infty} \hspace{-2.5mm} \Pr\(\frac{ G_{\text{BS}_i} h_{i} \ell(r)}{ I_{x_i} + \sigma^2} \geq T \hspace{-0.8mm} \mid \hspace{-0.8mm} S_i=1, r\) \hspace{-0.8mm} \Pr(S_i=1 \mid r) r\mathrm{d}r \nonumber\\
&= \Thetau \frac{\Thetab}{2\pi} \int_{0}^{\infty} \mathcal{L}_{I_{x_i}}^m (\frac{T r^{\alpha}}{G_{\text{BS}_i}}) e^{-T r^{\alpha} \sigma^2} e^{-\beta r} r\mathrm{d}r \:,
\end{align}

\subsection{Lemma~\ref{lem: sidelobe-success}}
Define $Z_k$ as the event that the SINR with sidelobe is larger than $T$ at mini-slot $k$ given distance $r = \lVert x_i \rVert$: $Z_k \triangleq \{\text{SINR}_{x_i}^k > T\}$. Unlike serving with the mainlobe, each BS has $N_c -1$ mini-slots sending pilot to the UE with sidelobe. Therefore, given a serving BS, both the desired signal and part of the interference (sidelobe part) are at same locations during $N_c -1$ mini-slots, which makes the events $Z_k$ and $Z_j$ not independent. In this case, we can consider the $N_c -1$ mini-slots as a single-input-multi-output (SIMO) system, i.e. one mini-slot as one receiving antenna. For better dealing with the correlated interference, we decompose the interference during $N_c -1$ mini-slots into two parts: correlated sidelobe interference with gain $\epsilon$ and independent mainlobe interference with gain $G_m^* = G_m - \epsilon$, where $G_m$ denotes the mainlobe gain of the antenna. 

We focus first on the probability of the joint occurrence of $Z_k$ over $n$ mini-slots, $P_n(T,r)$. Let $\delta(r) = \frac{T r^\alpha}{\epsilon}$ and $r = \lVert x_i \rVert$, we have:
\begin{align}
\label{equ:Pn}
P_{x_i}^n(T) &= \Pr\(\bigcap_{k \in [n]} Z_k\) = \Pr\(\text{SINR}_{x_i}^1 > T, \cdots, \text{SINR}_{x_i}^n > T\) \nonumber \\
&= \Pr\(h_1 > \delta(r)(I_1 + \sigma^2), \cdots, h_n > \delta(r)(I_n + \sigma^2)\) \nonumber \\
&= \mathbb{E}\[e^{-\delta(r)(I_1 + \sigma^2)} \cdots e^{-\delta(r)(I_n + \sigma^2)}\] \nonumber \\
&\stackrel{(a)}{=} \mathbb{E} \prod\limits_{k=1}^{n} e^{\delta(r)I_{km}}  \prod\limits_{k=1}^{n} e^{\delta(r)I_{ks}}  \prod\limits_{k=1}^{n} e^{\delta(r)\sigma^2}, \: 
\end{align}
where (a) follows from the decomposition of the interference into sidelobe part $I_{ks}$ and mainlobe part $I_{km}$.
The interferences caused by the sidelobe are correlated through the common randomness $\Phi$. Thus we obtain 
\begin{align}
\mathbb{E} \prod\limits_{k=1}^{n} e^{\delta(r)I_{ks}} &= \mathbb{E} \prod\limits_{k=1}^{n} \prod\limits_{\Phi_s} e^{\delta(r) \epsilon R_j^{-\alpha} h_{j,k} S_{j,k}} \nonumber \\
&\stackrel{(a)}{=} \mathbb{E} \prod\limits_{\Phi_s} \mathbb{E}_h \prod\limits_{k=1}^{n}  e^{\delta(r) \epsilon R_j^{-\alpha} h_{j,k}} e^{-\beta R_j} + 1 - e^{-\beta R_j}  \nonumber \\
&= \mathbb{E} \prod\limits_{\Phi_s} \( 1- \frac{T r^{\alpha} e^{-\beta t}}{t^{\alpha} + T r^{\alpha}}\)^n \nonumber \\
&= \exp \( -\Thetau \lambda \int_0^{\infty} \(1 - \( 1- \frac{T r^{\alpha} e^{-\beta t}}{t^{\alpha} + T r^{\alpha}}\)^n \) t \mathrm{d}t\) \nonumber \\
&\stackrel{(b)}{=} \exp \(- \frac{1}{2} \Thetau \lambda r^2 T^{\frac{2}{\alpha}}  \int_0^{\infty} \(1 - \( 1- \frac{e^{-\beta \sqrt{u} r T^{\frac{1}{\alpha}}}}{1 + u^{\frac{\alpha}{2}}}\)^n \)  \mathrm{d}u\),  \:
\end{align}
where (a) follows from the independent fading assumption and (b) is derived by employing a change of variable $u=\(\frac{t}{rT^{\frac{1}{\alpha}}}\)^2$. In contrast, the interferences caused by the mainlobe are independent. Thus we obtain
 
\begin{align}
\mathbb{E} \prod\limits_{k=1}^{n} e^{\delta(r)I_{km}} &= \mathbb{E} \prod\limits_{k=1}^{n} \prod\limits_{\Phi_m} e^{\delta(r) G_{m}^* R_j^{-\alpha} h_{j,k} S_{j,k}} \nonumber \\
&\stackrel{(a)}{=} \prod\limits_{k=1}^{n} \mathbb{E} \prod\limits_{\Phi_m} \mathbb{E}_h   e^{\delta(r)P_{ts} R_j^{-\alpha} h_{x_j}} e^{-\beta R_j} + 1 - e^{-\beta R_j}  \nonumber \\
&= \prod\limits_{k=1}^{n} \mathbb{E} \prod\limits_{\Phi_m} \( 1- \frac{\delta(r) G_m^* R_j^{-\alpha} e^{-\beta R_j}}{1 + \delta(r) G_m^* R_j^{-\alpha}}\) \nonumber \\
&= \prod\limits_{k=1}^{n} \exp \( -\Thetau \frac{\Thetab}{2\pi} \lambda \int_0^{\infty} \frac{T G_m^* r^{\alpha} e^{-\beta t}}{\epsilon t^{\alpha} + T G_m^* r^{\alpha}} t \mathrm{d}t\) \nonumber \\
&\stackrel{(b)}{=} \exp \( -n \Thetau \frac{\Thetab}{2\pi} \lambda \int_0^{\infty} \frac{T G_m^* r^{\alpha} e^{-\beta t}}{\epsilon t^{\alpha} + T G_m^* r^{\alpha}} t \mathrm{d}t\),  \:
\end{align}
where (a) follows from the independent interference among mini-slots. 

In the next step, we need to derive the probability $Q_{x_i}^n(T)$ that the SINR in at least one mini-slot exceeds the threshold. This can be viewed as a selection combining in SIMO system where a successful transmission occurs if $\max_{k \in [n]} \{\text{SINR}^k > T\}$. Therefore, $Q_{x_i}^n(T)$ can be expressed as:

\begin{align}\label{eq:selection}
Q_{x_i}^n(T) = \Pr\(\bigcup_{k=1}^{n} Z_k \) = \sum\limits_{k=1}^{n}(-1)^{k+1} \binom{n}{k} P_{x_i}^k(T)
\end{align}

Finally, the probability of successful detection with the sidelobe over $n$ mini-slots $P_{ss}$ is the selection combining from $\Phi_s$. Define event $V_{x_i} \triangleq \{Q_{x_i}^n(T)\}$, then $P_{ss}$ can be expressed as:
\begin{align}
\label{equ:Pss1}
P_{\text{ss}} &= \Pr\(\bigcup_{x_i \in \Phi_s} V_{x_i} \) \nonumber \\
&= \frac{2\pi - \Thetab}{2\pi} \lambda \int_{\mathbb{R}^2} Q_{x_i}^n(T) \mathrm{d}x_i \nonumber\\
&= \Thetau \frac{2\pi - \Thetab}{2\pi} \lambda \int_{0}^{\infty} Q_{x_i}^n(T)  e^{-\beta r} r\mathrm{d}r \:,
\end{align}

\subsection{Proposition~\ref{prop: detection-failure-prob-sidelobe}}

Since we assume the mainlobe and sidelobe detections are independent, the failure of them are independent as well. Thus, the detection failure probability can be written as:
\begin{equation}\label{eq: Pfsidelobeproof}
P_f^{\text{S}} = (1- P_{ss})(1- P_{sm})^{N_c} \:.
\end{equation}

\subsection{Proposition~\ref{prop: expected-latency}}
We assume that the UE and BSs are independent over different frames. The reason is twofold: firstly, we can consider mobility scenario where UEs or obstacles are moving, similar as in \cite{Li2017a}; secondly, the UE may not search over all the potential directions in one frame and will turn to another direction in the next frame. Thus even for static UE, we can consider the UE is pointing to different BS in different frames. Therefore, we define $M \in \mathbb{N}^{+}$ as the number of frames for the typical UE to detect a BS and $M$ follows a geometric distribution with parameter $1-P_f$. The IA latency is then the sum of failed frames and one IA period as: 
\begin{align}
\mathbb{E}[D_I(N_c)] = \(\frac{1}{1-P_f(N_c)} - 1\)T_{f} + T_{\text{cs}} + T_{\text{ra}}\:.
\end{align}

At last, the total transmission latency is comprised of three parts: the failed frame duration, the IA period and the data transmission period as below:
\begin{equation}
    \mathbb{E}[D_T(N_c)] = \(\frac{1}{1-P_f(N_c)} - 1\)T_{f} + \left\lceil \frac{L}{R_T(T_f- T_{\text{cs}} - T_{\text{ra}})} \right\rceil (T_{\text{cs}} + T_{\text{ra}}) + \frac{L}{R_T} \label{eq: datalatencyproof}\:.
\end{equation}



\bibliographystyle{./Components/MetaFiles/IEEEtran}
\bibliography{./Components/MetaFiles/References}

\end{document}